 \documentclass{aa}  

\usepackage{graphicx}
\usepackage{txfonts}
\usepackage{soul}

% ...
\begin{document}


%


    \title{Statistical study of electron density turbulence and ion-cyclotron waves in the inner heliosphere: Solar Orbiter observations}

	\titlerunning{Turbulence of solar wind density fluctuations: Solar Orbiter observations}
	
	
	\author{F. Carbone\inst{1}
		\and L. Sorriso-Valvo\inst{2,3}\fnmsep\thanks{Corresponding author: lucasorriso@gmail.com}
		\and Yu. V. Khotyaintsev\inst{2}
		\and K. Steinvall\inst{2}
		\and A. Vecchio\inst{4,5}
		\and D. Telloni\inst{6}
		\and E. Yordanova\inst{2}
		\and D. B. Graham\inst{2}
		\and N. J. T. Edberg\inst{2}
		\and A. I. Eriksson\inst{2}
		\and E. P. G. Johansson\inst{2}
		\and C. L. V\'asconez\inst{7}
		\and M. Maksimovic\inst{5} 
		\and R. Bruno\inst{8}
		\and R. D'Amicis\inst{8}
		\and S. D. Bale\inst{9,10}
		\and T. Chust\inst{11} 
		\and V. Krasnoselskikh\inst{12} 
		\and M. Kretzschmar\inst{12,13}
		\and E. Lorf\`evre\inst{14} 
		\and D. Plettemeier\inst{15}
		\and J. Sou\v{c}ek\inst{16} 
		\and M. Steller\inst{17}
		\and \v{S}. \v{S}tver\'ak\inst{18,16} 
		\and P. Tr\'avn\'i\v{c}ek\inst{9,16}
		\and A. Vaivads\inst{12,19} 
		\and T. S. Horbury\inst{20}
		\and H. O'Brien\inst{20}
		\and V. Angelini\inst{20}
		\and V. Evans\inst{20}
	}
	
	\institute{
		National Research Council -- Institute of Atmospheric Pollution Research, C/o University of Calabria, 87036 Rende, Italy
		\and Swedish Institute of Space Physics (IRF), \r{A}ngstr\"om Laboratory, Lägerhyddsvägen 1, 75121 Uppsala, Sweden
		\and CNR, Istituto per la Scienza e Tecnologia dei Plasmi, via Amendola 122/D, 70126, Bari, Italy
		\and LESIA, Observatoire de Paris, Universit\'e PSL, CNRS, Sorbonne, Universit\'e, Univ. Paris Diderot, Sorbonne Paris Cit\'e, 5 place Jules Janssen, 92195 Meudon, France
		\and Research Institute for Mathematics, Astrophysics and Particle Physics, Radboud University, Nijmegen, The Netherlands  
		\and National Institute for Astrophysics - Astrophysical Observatory of Torino, Italy
		\and Departamento de F\'isica, Escuela Polit\'ecnica Nacional, Ladr\'on de Guevara E11-253, 170525, Quito, Ecuador 
		\and National Institute for Astrophysics (INAF) - Institute for Space Astrophysics and Planetology (IAPS), Via Fosso del Cavaliere, 100, 00133 Rome, Italy
		\and Space Sciences Laboratory, University of California, Berkeley, CA, USA
		\and Physics Department, University of California, Berkeley, CA, USA
        \and LPP, CNRS, Ecole Polytechnique, Sorbonne Universit\'{e}, Observatoire de Paris, Universit\'{e} Paris-Saclay, Palaiseau, Paris, France
		\and LPC2E, CNRS, 3A avenue de la Recherche Scientifique, Orl\'eans, France
		\and Universit\'e d'Orl\'eans, Orl\'eans, France
		\and CNES, 18 Avenue Edouard Belin, 31400 Toulouse, France
		\and Technische Universit\"at Dresden, Würzburger Str. 35, D-01187 Dresden, Germany
		\and Institute of Atmospheric Physics of the Czech Academy of Sciences, Prague, Czechia
		\and Space Research Institute, Austrian Academy of Sciences, Graz, Austria
		\and Astronomical Institute of the Czech Academy of Sciences, Prague, Czechia
		\and Department of Space and Plasma Physics, School of Electrical Engineering and Computer Science, Royal Institute of Technology, Stockholm, Sweden;
		\and Space and Atmospheric Physics, The Blackett Laboratory, Imperial College of London, London, SW7 2AZ, UK}
	
	\authorrunning{F. Carbone et al.}
	
	\date{\today}
	
	
	\abstract
	{The recently released spacecraft potential measured by the RPW instrument on-board Solar Orbiter has been used to estimate the solar wind electron density in the inner heliosphere.}
	{Solar-wind electron density measured during June 2020 has been analysed to obtain a thorough characterization of the turbulence and intermittency properties of the fluctuations. Magnetic field data have been used to describe the presence of ion-scale waves.}
	{Selected intervals have been extracted to study and quantify the properties of turbulence. The Empirical Mode Decomposition has been used to obtain the generalized marginal Hilbert spectrum, equivalent to the structure functions analysis, and additionally reducing issues typical of non-stationary, short time series. The presence of waves was quantitatively determined introducing a parameter describing the time-dependent, frequency-filtered wave power.}
	{A well defined inertial range with power-law scaling has been found almost everywhere. However, the Kolmogorov scaling and the typical intermittency effects are only present in part of the samples. Other intervals have shallower spectra and more irregular intermittency, not described by models of turbulence. 
	These are observed predominantly during intervals of enhanced ion frequency wave activity. Comparisons with compressible magnetic field intermittency (from the MAG instrument) and with an estimate of the solar wind velocity (using electric and magnetic field) are also provided to give general context and help determine the cause for the anomalous fluctuations.}
	{}
	
	
	\maketitle

	%
	\section{Introduction}

The turbulent nature of solar wind fluctuations has been investigated for more than half a century~\citep[see, e.g.][]{Bruno2016}. 
Advances are constantly achieved thanks to the increasingly accurate measurements of several dedicated space mission, to the enormous improvement of numerical calculation, to new detailed models and theoretical frameworks, and to the development of specific data analysis techniques. Nevertheless, the extremely complex nature of the system and the coexistence of multiple actors, scales, and dynamical regimes result in a large number of still open questions~\citep{Viall2020}. Among these, the very nature of the turbulent cascade of the solar wind flow and its relationship with the small-scale processes still need to be described in full~\citep{tu1995,Bruno2013,Matthaeus2011,chen_2016}. 
Magnetic field fluctuations have been characterized with great detail at magnetohydrodynamic and kinetic scales, for example through spectral and high-order moments analysis~\citep[][]{tu1995,Bruno2013}. The anisotropic nature of magnetic turbulence has also been addressed, and is still being debated, due to the limited access to three-dimensional measurements in space~\citep[see, e.g.][]{horbury2008,Sorriso2010,horbury2012,Yordanova2015,verdini2018,Telloni2019,Oughton2020}. 
Velocity fluctuations have been studied thoroughly~\citep[see, e.g.][]{Sorrisovalvo1999,Bruno2013}, although the kinetic scales still remain quite unexplored for instrumental limitations, most notably in sampling time resolution.
Both velocity and magnetic field show highly variable turbulence properties, with well developed spectra, strong intermittency~\citep[][]{Sorrisovalvo1999}, anisotropy, and linear third-order moments scaling~\citep[][]{Sorriso2007,Carbone2011}.
The level of Alfv\'enic fluctuations~\citep[mostly but not exclusively found in fast streams, see e.g.][]{DAmicis2011,Bruno2019} are believed to be associated with the state of the turbulence. 
In particular, solar wind samples containing more Alfv\'enic fluctuations are typically associated with less developed turbulence, as inferred from both shallower spectra and reduced intermittency~\citep[see][and references therein]{Bruno2013}. This is consistent with the expectation that uncorrelated Alfv\'enic fluctuations contribute to reduce the nonlinear cascade by sweeping away the interacting structures~\citep{Dobrowolny1980}, as also confirmed by the observed anticorrelation between the turbulent energy cascade rate and the cross-helicity~\citep{Smith2009,Marino2011,Marino2011b}.

Conversely, density fluctuations have been only partially explored, due in part to the unavailability of high-frequency time series, and in part to their supposedly secondary relevance in the nearly incompressible solar-wind dynamics. 
In recent years, studies have shown that proton density is also turbulent and intermittent~\citep{Hnat2003,Hnat2005}, with the turbulence characteristics often depending of the Alfv\'enic nature of each specific solar wind interval. In particular, it is understood that in the nearly incompressible Alfv\'enic solar wind the turbulence of density fluctuations is strongly similar to that of magnetic field magnitude, both behaving like scalar quantities passively advected by velocity and magnetic components turbulence~\citep{Goldreich1995,Chen2012}. 
Conversely, in the more compressive non-Alfv\'enic solar wind (more typically associated with slow streams) they are actively contributing to the nonlinear transfer of energy~\citep{Schekochihin2009,Boldyrev2013}. 
The radial evolution of proton density turbulence was examined, providing evidence of complex, unexpected behaviours~\citep{Bruno2014}. Modeling of the radial evolution was attempted based on the parametric instability, expected to generate increasingly compressive fluctuations as the solar wind expands away from the sun. However, despite the analysis provided some insight, the lack of statistical description still prevents the validation of models that could provide a prediction for the radial evolution of the density fluctuations. 
Finally, sub-ion scale density turbulence has been performed using high-resolution proton density measurements from Spektr-R~\citep{Chen2014,Riazantseva2019} and electron density from MMS~\citep{Roberts2020}, but results could not be fully conclusive about the nature of the multifractal and intermittency properties of the fluctuations~\citep{Sorrisovalvo2017,Carbone2018}. 
Therefore, a deeper analysis of solar wind density fluctuations is necessary in order to constrain modelings and to understand the role of density fluctuations in the magnetohydrodynamic turbulent cascade. Note that, although the above studies may refer either to proton or electron density, depending on the instrument used for the measurements, at the scales of interest for turbulence the two can be safely considered equal for the quasi neutrality condition of solar wind plasma.

Recent development in electromagnetic wave payload has allowed to obtain density fluctuations from the measurement of the spacecraft potential~\citep{Pedersen1995,roberts2017b}. This gave access to higher-frequency, accurate density measurements.
In particular, the Solar Orbiter spacecraft~\citep{Muller2020} was launched in February 2020, equipped with both remote and in situ instrumentation aiming at investigating the sun and the solar wind from and within the inner heliospehre~\citep{Zouganelis2020}. 
The Radio and Plasma Waves (RPW) instrument~\citep{Maksimovic2020} measures the spacecraft potential with unprecedented accuracy, allowing the estimation of the high-cadence solar wind electron density~\citep{Yuri2021}. 
Together with the enhanced data quality provided by Solar Orbiter instruments, novel data analysis techniques are emerging that improve the delicate measurement of turbulence parameters in solar wind data. Among these, the Empirical Mode Decomposition~\citep[EMD,][]{Huang2008} has been successfully used to mitigate short-sample and large-scale structures effects~\citep{Carbone2018,Tommasi2019}. 
This paper aims at providing the first description of the properties of turbulence as obtained applying EMD-based analysis techniques to the Solar Orbiter RPW density measurements taken during the month of June 2020~\citep{Yuri2021}. 

The results are discussed in relation to solar wind parameters and magnetic field turbulence, studied applying the same EMD-based techniques to the MAG instrument~\citep{Horbury2020}. 
Moreover, after introducing a novel parameter to describe the presence of ion-scale waves, the relationship between the properties of turbulence and such waves has been analyzed.

Section~\ref{sec:data} provides a description of the data used for the analysis. 
In Section~\ref{sec:analysis} the techniques used for the analysis of turbulent fluctuations and the corresponding results are described. 
In Section~\ref{sec:waves} the wave parameter is introduced and the presence of ion-scale waves is discussed. 
Section~\ref{sec:discussion} gives a detailed discussion on the existing correlations between solar wind, turbulence and wave parameters and their implications. 
Finally, conclusions are summarized in Section~\ref{sec:conclusions}.

	\section{Description of data}
	\label{sec:data}

In order to study the properties of turbulence of solar wind density fluctuations, we make use of Solar Orbiter measurements taken from 7 to 29 June 2020, when the spacecraft was orbiting the Sun between 0.52 AU and 0.55 AU. During that time interval, the Solar Wind Analyser (SWA) plasma instruments~\citep{Owen2020} were not operational, so that direct measurements of proton and electron moments are not available. Here, we use the 16 Hz electron density $n_e$, accurately estimated from the RPW probe-to-spacecraft potential measurements~\citep{Maksimovic2020}.
The equilibrium floating potential of the spacecraft is reached when the current due to photo-electrons emitted from the spacecraft is balanced by the plasma current collected by the spacecraft. The equilibrium is reached instantaneously (0.1-1 ms) on the time scales of interest for the turbulence studies. From the current balance we can find that the electron density has an exponential relation to the spacecraft potential. Then making an exponential fit of the spacecraft potential to the electron density obtained from the high-frequency measurement of plasma quasi-thermal noise one can find the relation for approximate conversion of the probe-to-spacecraft potential to electron density. Details on the density estimation technique are given in~\citet{Yuri2021}. 

Localized estimates of the solar wind speed from the deHoffmann-Teller (HT) analysis of electromagnetic fluctuations are used to provide general context, most notably discriminating between fast and slow solar wind streams.
HT analysis is used to find the velocity of the frame in which the electric field is zero. In the solar wind, where current sheets and MHD turbulence are ubiquitous, the HT velocity is in general a good estimate of the solar wind velocity. In order to get reasonable coverage, estimates of the solar wind speed $V_{sw}$ were obtained every ten minutes by applying the HT analysis on one-hour running intervals of E and B data. Details about the HT analysis can be found in~\citet{Steinvall2021}.

Finally, magnetic field vector $\mathbf{B}$ measurements from the magnetometer (MAG)~\citep{Horbury2020} are studied for comparison. 

Due to the presence of large gaps in the electron density data, and to avoid strong violation of homogeneity and stationarity, a number of intervals of variable length were extracted from the electron density time series $n_e(t)$. All intervals were chosen as relatively stationary, covering at least $1$ hour data, and could only include a few data gaps shorter than $5$ seconds. The remaining missing points have been interpolated linearly. 
Such criteria resulted in the selection of $36$ sub-intervals, whose list and macroscopic details (starting and ending time, interval duration, mean distance from the Sun, estimated mean solar wind bulk speed, and the angle between the ambient magnetic field and the radial direction averaged over each interval) are given in Table~\ref{table1:data_params}. 
Given the estimated solar wind speed, the Taylor hypothesis~\citep{Taylor1938} is considered valid throughout the whole data set, and allows to interpret the time series as equivalent to a space ensemble, therefore enabling the standard tools for turbulence analysis.

\begin{figure*}
		\centering\includegraphics[scale=0.35]{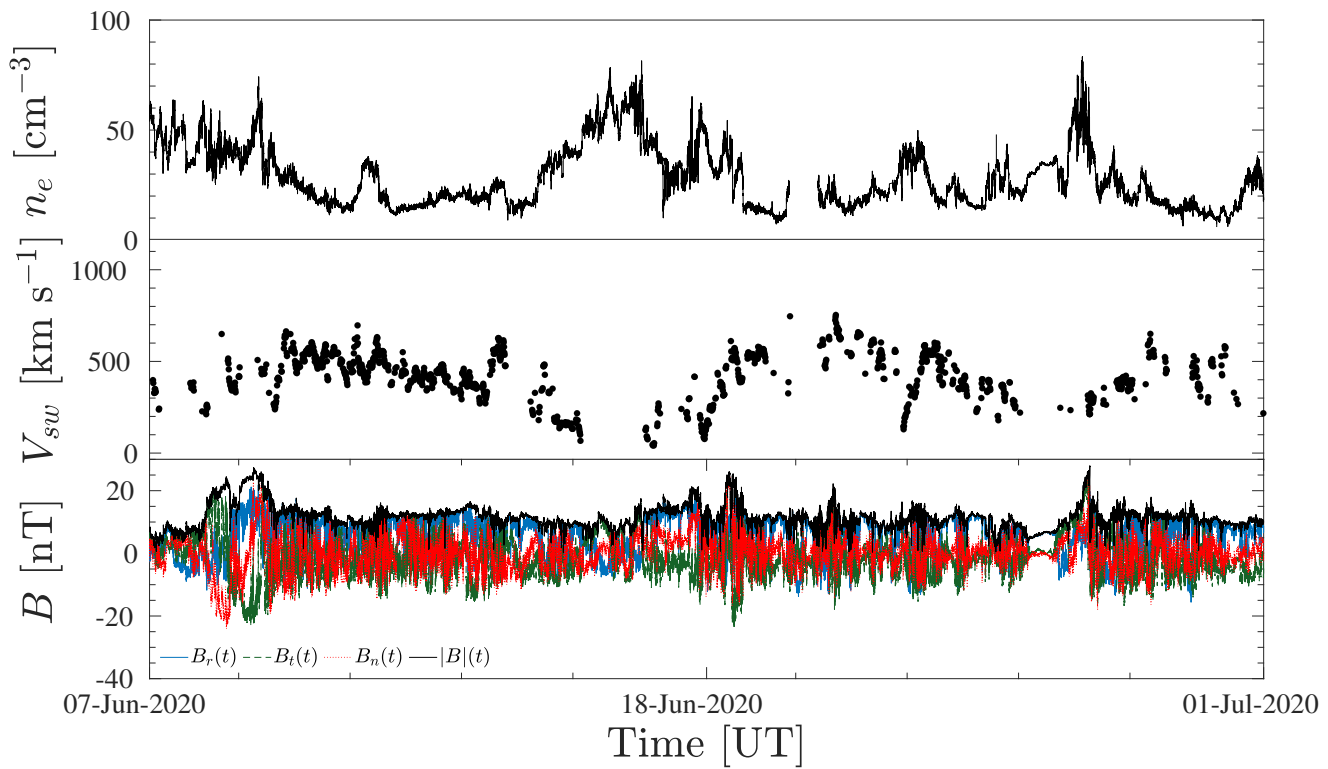}
		\caption{From top to bottom: solar wind electron density $n_e$ (RPW), deHoffmann-Teller solar wind velocity estimate $V_{sw}$ (RPW) and interplanetary magnetic field components $B_i$ and magnitude $|B|$ (MAG), measured by Solar Orbiter during the whole month of June 2020.}
		\label{fig:overview}
	\end{figure*}

Figure~\ref{fig:overview} shows an overview of the above parameters for the whole month of June 2020. 
The overall solar wind conditions are variable, with alternation of fast and slow streams, as well as complex coronal mass ejection structures~\citep[as studied in detail by][]{Telloni2021}. However, the selected sub-intervals, whose duration is between one and six hours, are typically embedded in homogeneous solar wind conditions.
%

	\begin{table*}
		\caption{List of $36$ intervals selected for this work. For each interval, the list includes initial and final time, duration $\Delta \mathcal{W}$, estimated deHoffmann-Teller solar wind speed $V_{sw}$, average angle between the magnetic field and the radial direction $\theta_{vb}$, and the identified group (see Section~\ref{sec:groups}).}             
		\label{table1:data_params}      
		\centering                          
		\begin{tabular}{c c c c c c c}        
			\hline\hline                 
			Sample & Start time & End time & $\Delta \mathcal{W}$ $[$hour$]$ & $V_{sw}$ $[$km/s$]$ & $\theta_{vb}$ & Group \\    
			\hline 			
			\multicolumn{7}{c}{07-June-2020} \\
			\hline
1  & 05:22:13 & 06:28:53 & 1.11 & 228        & 80  &  1  \\
2  & 09:26:39 & 12:02:13 & 2.59 & 353.84     & 95  &  1  \\ 
3  & 12:35:33 & 13:48:47 & 1.22 & 353.84     & 78  &  1  \\	
4  & 14:00:00 & 16:39:26 & 2.66 & 650        & 86  &  1  \\
5  & 16:42:33 & 19:57:46 & 3.25 & 430$\pm$62 & 88  &  1  \\
6  & 20:05:53 & 23:47:59 & 3.70 & 354$\pm$12 & 128  & 1  \\	
			\hline  
			\multicolumn{7}{c}{08-June-2020} \\
			\hline
7  & 00:00:00 & 01:51:06 & 1.85 & 430$\pm$25 & 130  & 1 \\
8  & 01:56:39 & 03:52:23 & 1.93 & 398.07     & 117  & 2 \\
9  & 04:00:00 & 07:08:53 & 3.15 & 415.76     & 127  & 2 \\
10 & 07:31:06 & 10:28:53 & 2.96 & 500        & 129  & 2 \\
11 & 10:39:59 & 11:53:11 & 1.22 & 415.76     & 123  & 1 \\
12 & 13:35:00 & 15:52:59 & 2.30 & 440$\pm$12 & 131  & 2 \\
			\hline  
			\multicolumn{7}{c}{09-June-2020} \\
			\hline
13 & 00:00:00 & 03:57:59 & 3.97 & 614$\pm$43 & 133  & 3 \\
14 & 04:10:33 & 10:01:59 & 5.86 & 514$\pm$50 & 157  & 1 \\
15 & 10:59:59 & 15:58:59 & 4.98 & 555$\pm$27 & 148  & 1 \\
16 & 18:02:46 & 21:00:33 & 2.96 & 485$\pm$16 & 143  & 3 \\
17 & 21:01:40 & 23:52:46 & 2.85 & 535$\pm$47 & 145  & 1 \\
			\hline  
			\multicolumn{7}{c}{10-June-2020} \\
			\hline
18 & 04:05:17 & 07:36:23 & 3.52 & 522$\pm$25 & 155  &  2 \\
19 & 07:47:30 & 12:58:37 & 5.19 & 443$\pm$35 & 136  &  2 \\
20 & 19:00:50 & 22:09:44 & 3.15 & 425$\pm$25 & 149  &  2 \\
			\hline  
			\multicolumn{7}{c}{11-June-2020} \\
			\hline
21 & 00:00:00 & 02:57:46 & 2.96 & 541$\pm$55 & 154  &  2\\
22 & 04:05:17 & 06:57:30 & 2.87 & 529$\pm$27 & 159  & 1 \\
23 & 09:38:37 & 12:36:24 & 2.96 & 397$\pm$37 & 138  & 3 \\
24 & 17:00:50 & 21:27:30 & 4.44 & 407$\pm$42 & 118  & 1 \\
			\hline  
			\multicolumn{7}{c}{14-June-2020} \\
			\hline
25 & 00:00:00 & 01:01:06 & 1.02 & 415.76     & 149  & 2 \\
			\hline  
			\multicolumn{7}{c}{20-June-2020} \\
			\hline	
26 & 00:55:33 & 03:08:53 & 2.22 & 575 (e)    & 172  & 1 \\
27 & 03:19:59 & 06:55:33 & 3.59 & 525 (e)    & 165  & 2 \\
			\hline  
			\multicolumn{7}{c}{22-June-2020} \\
			\hline	
28 & 01:58:09 & 04:33:42 & 2.59 & 542$\pm$75 & 118  & 1 \\
29 & 16:08:47 & 18:54:38 & 2.76 & 517$\pm$59 & 120  & 1 \\
			\hline  
			\multicolumn{7}{c}{24-June-2020} \\
			\hline	
30 & 04:26:39 & 06:17:46 & 1.85 & 379$\pm$16 & 176  & 2 \\   
31 & 14:30:33 & 16:06:39 & 1.60 & 300        & 169  & 3 \\  
32 & 16:47:46 & 17:57:46 & 1.17 & 403        & 147  & 1 \\
			\hline  
			\multicolumn{7}{c}{27-June-2020} \\
			\hline	
33 & 05:33:19 & 07:57:46 & 2.41 & 283.07     & 79  &  1 \\
34 & 08:19:59 & 11:06:39 & 2.78 & 303$\pm$48 & 61  &  1 \\
35 & 19:02:01 & 22:55:21 & 3.89 & 386$\pm$28 & 118  & 1 \\	
			\hline  
			\multicolumn{7}{c}{29-June-2020} \\
			\hline	
36 & 16:20:33 & 18:00:33 & 1.67 & 278        & 162  & 2 \\
			\hline
		\end{tabular}
	\end{table*}


	\section{Analysis of solar wind electron density turbulence}
	\label{sec:analysis}

\begin{figure*}
		\centering\includegraphics[scale=0.45]{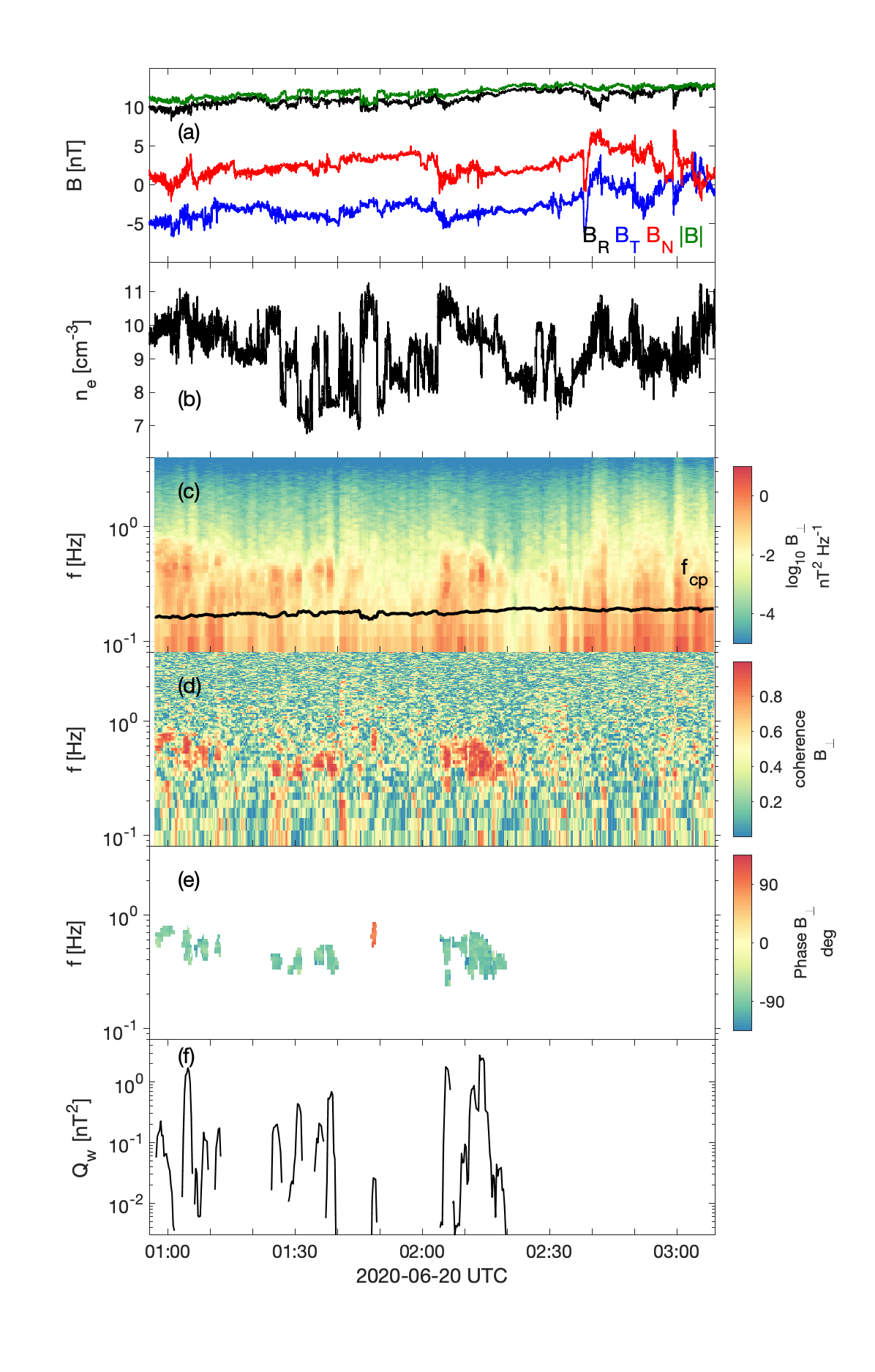}
		\centering\includegraphics[scale=0.45]{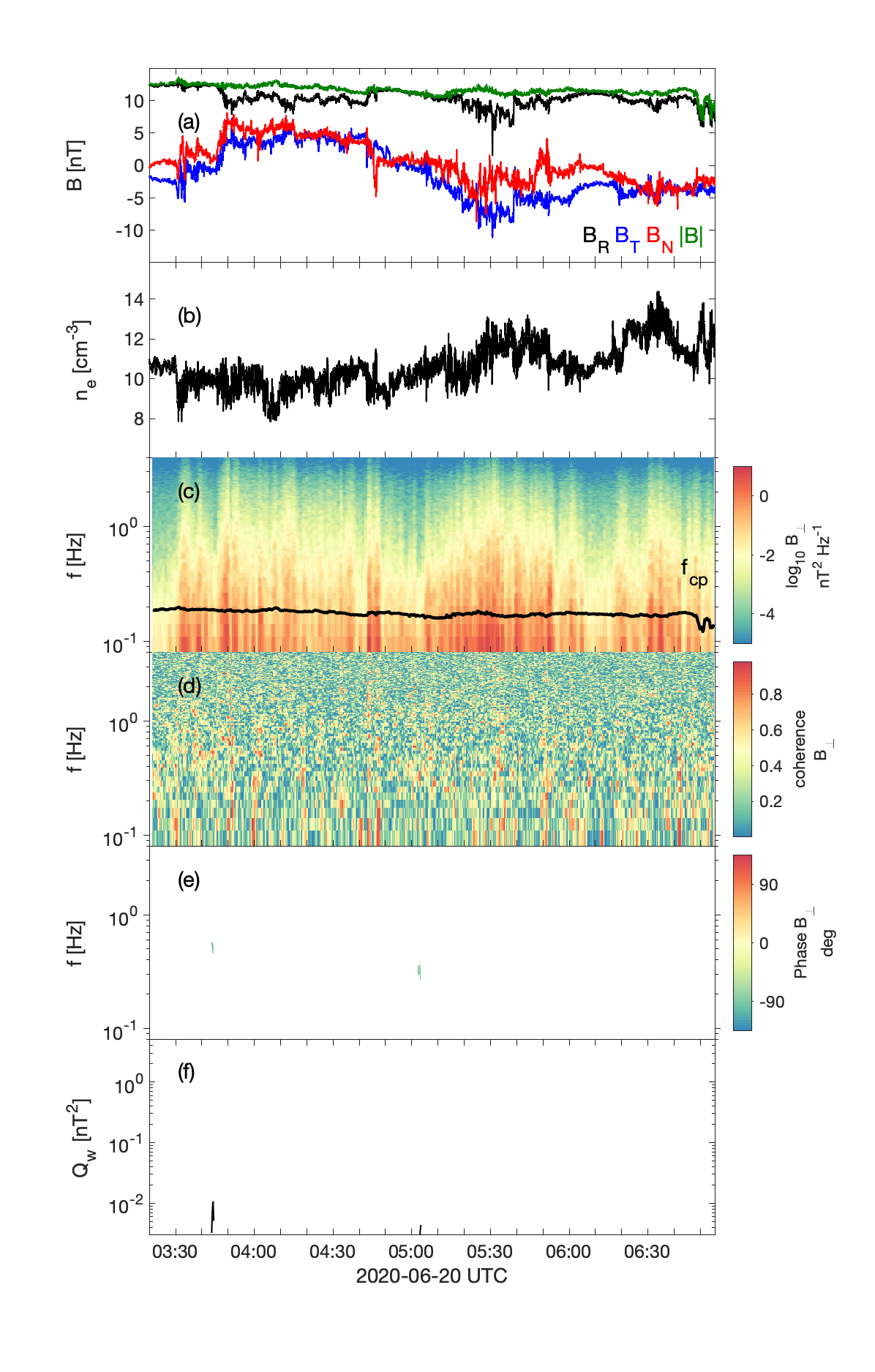}
		\caption{Two examples of adjacent intervals during day 20 June 2020. Left: interval 26; right panel: interval 27. From top to bottom: interplanetary magnetic field components $B_i$ and magnitude $|B|$ (MAG); electron density $n_e$ (RPW); magnetic filed spectrogram; perpendicular magnetic field components coherence; perpendicular magnetic field components phase angle; wave parameter $Q_w(t)$ (see Section~\ref{sec:waves}).}
		\label{fig:zoom}
	\end{figure*}

The properties of turbulent fields are usually studied through standard statistical analysis techniques. Among others, these may include~\citep{DudokDeWit2013}: the autocorrelation function, used to determine specific scales of the data; the power spectral density $E(f)$, providing information on the self-similar energy redistribution among scales; and the structure functions (SF) $S_q(\ell_t) \equiv \langle [n_e(t+\ell_t) - n_e(t)]^q \rangle \sim \ell_t^{\zeta(q)}$, with $\ell_t$ representing the time scale of the field increments, whose anomalous scaling exponents $\zeta(q)$ are able to quantitatively describe the effects of intermittency, namely the inhomogeneous nature of the turbulent cascade~\citep{Frish1995}. Related to the structure functions, the kurtosis of the distribution function of the fluctuations $K(\ell_t)=S_4(\ell_t) / S_2^2(\ell_t)$ is often used to quantify the intermittency effects. 
However, such techniques may be sensitive to the data sample characteristics, resulting in undesired effects not attributed to the turbulent energy cascade, but rather due, for example, to finite sample size, limited stationarity, or presence of superposed structures larger than the typical turbulence scales (for example non-turbulence related current sheets or velocity shears). Such issues are often occurring in ecliptic solar wind intervals, where instrumental performance and the intrinsic wind variability may prevent ideal experimental conditions for the study of turbulence~\citep{Carbone2018}. 
The magnification shown in the top two panels Figure~\ref{fig:zoom} shows an example of selected intervals (specifically, sub-intervals 26 and 27, both on June 20). 
The presence of large-scale modulation of the density profile, sporadic sharp gradients similar to ramp-cliff structures, and the general lack of strict stationarity appear evident~\citep{Mattheaus1983,Perri2010}. 

In order to mitigate the effects of such large-scale features, we make use of the Hilbert-Huang transform to obtain more precise estimators of the generalized high-order spectra.  
%
%
%

	\subsection{Empirical Mode Decomposition}
	\label{sec:hurst}

The technique used here is based on the Empirical Mode Decomposition (EMD)~\citep{Huang1998,Janosi2005,Carbone_emd2016}. This is a self-consistent, data-driven projection of a time series (as in this case the solar wind electron density $n_e(t)$) on a finite number $n$ of empirical basis functions $\phi_j(t)$, called intrinsic mode functions (IMFs), so that 
	\begin{equation}
	n_e(t) = \sum_{j=1}^n \phi_j(t) + r_n(t) \ .
	\label{eq1:emd}
	\end{equation}
The additive residual function $r_n(t)$ describes the mean trend. Each IMF can be characterized by the instantaneous timescale $\tau_j(t)$.
The decomposition is based on a recursive procedure, consisting of two main stages~\citep{Huang1998}: (i) the local extrema of $n_e(t)$ are interpolated through cubic spline to provide superior and inferior envelops of the time series; and (ii) the average between the two envelops is subtracted from the original data. The resulting field is accepted as an IMF if it satisfies the following specific criteria: the number of local extrema and zero crossings does not differ by more than one, and the average between the IMF superior and inferior envelopes is zero at all times. Otherwise, the procedure is repeated on the remaining IMF until the criteria are met according to the so called 3-thresholds stoppage criterion~\citep[with the following standard choice of parameters: $\delta = 0.05$, $\xi_1 = 0.05$, and $\xi_2 = 10\xi_1$, see][]{Rilling2003}.
	
\begin{figure*}
		\centering\includegraphics[scale=0.33]{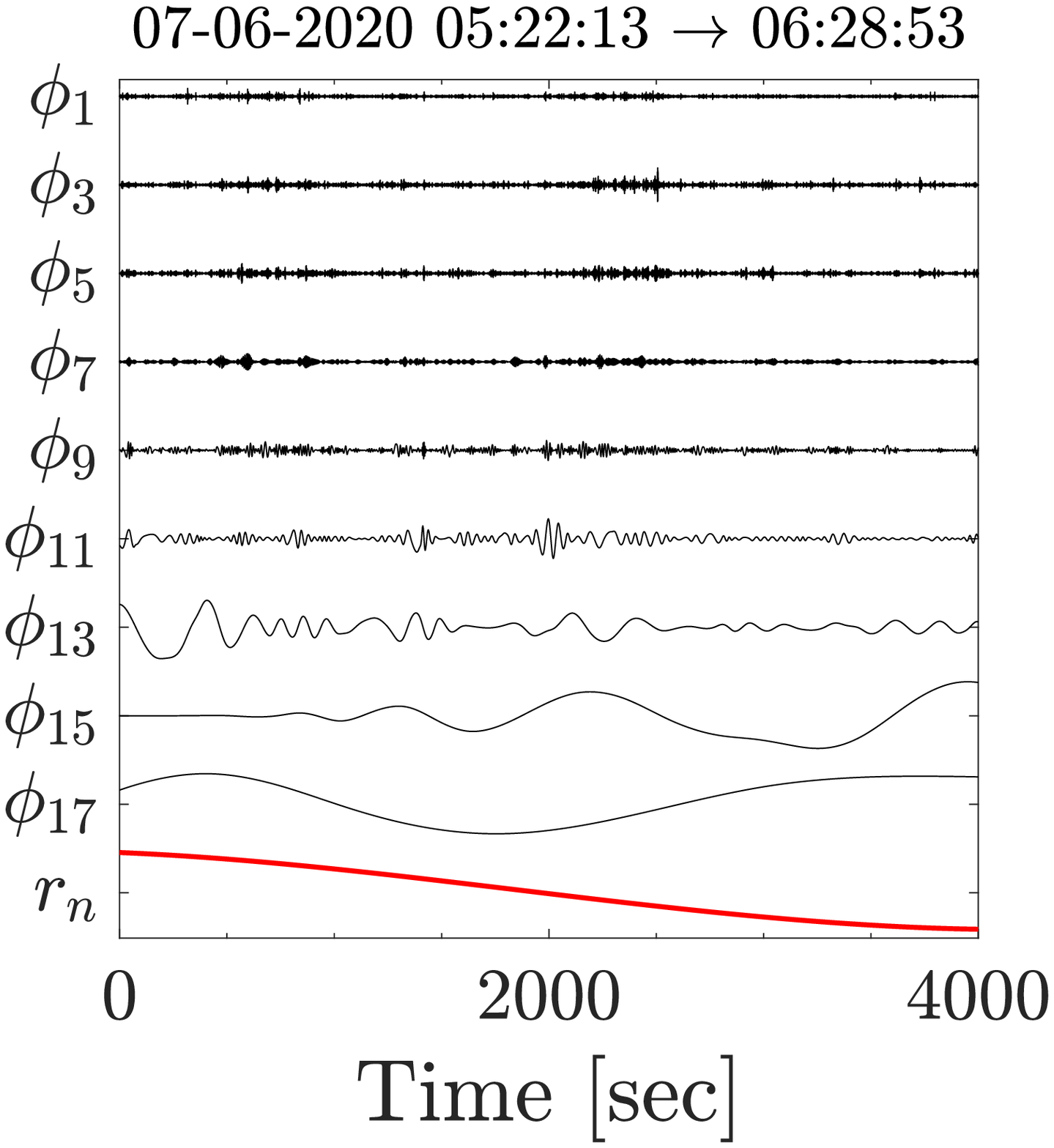}\includegraphics[scale=0.33]{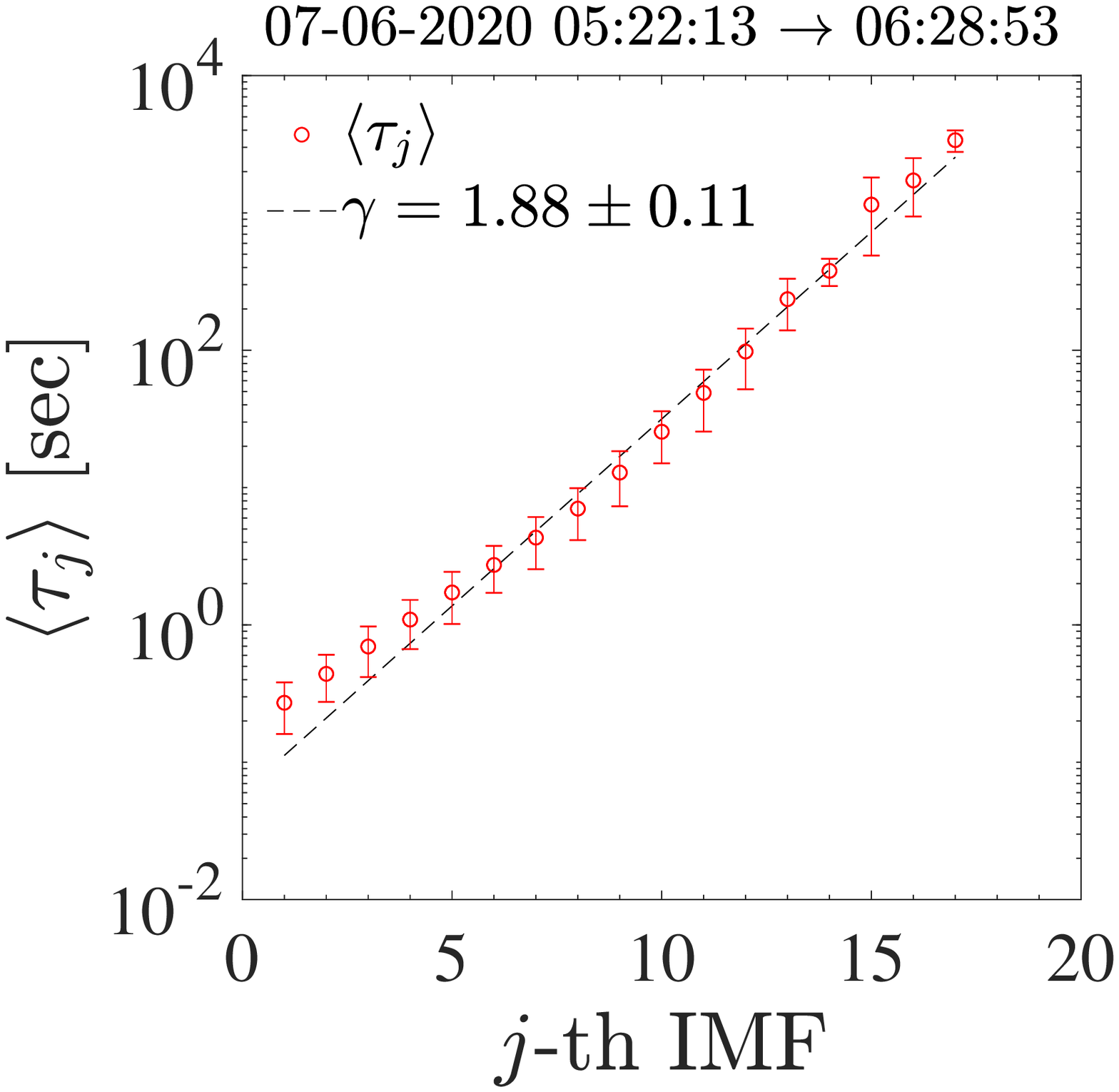}\includegraphics[scale=0.33]{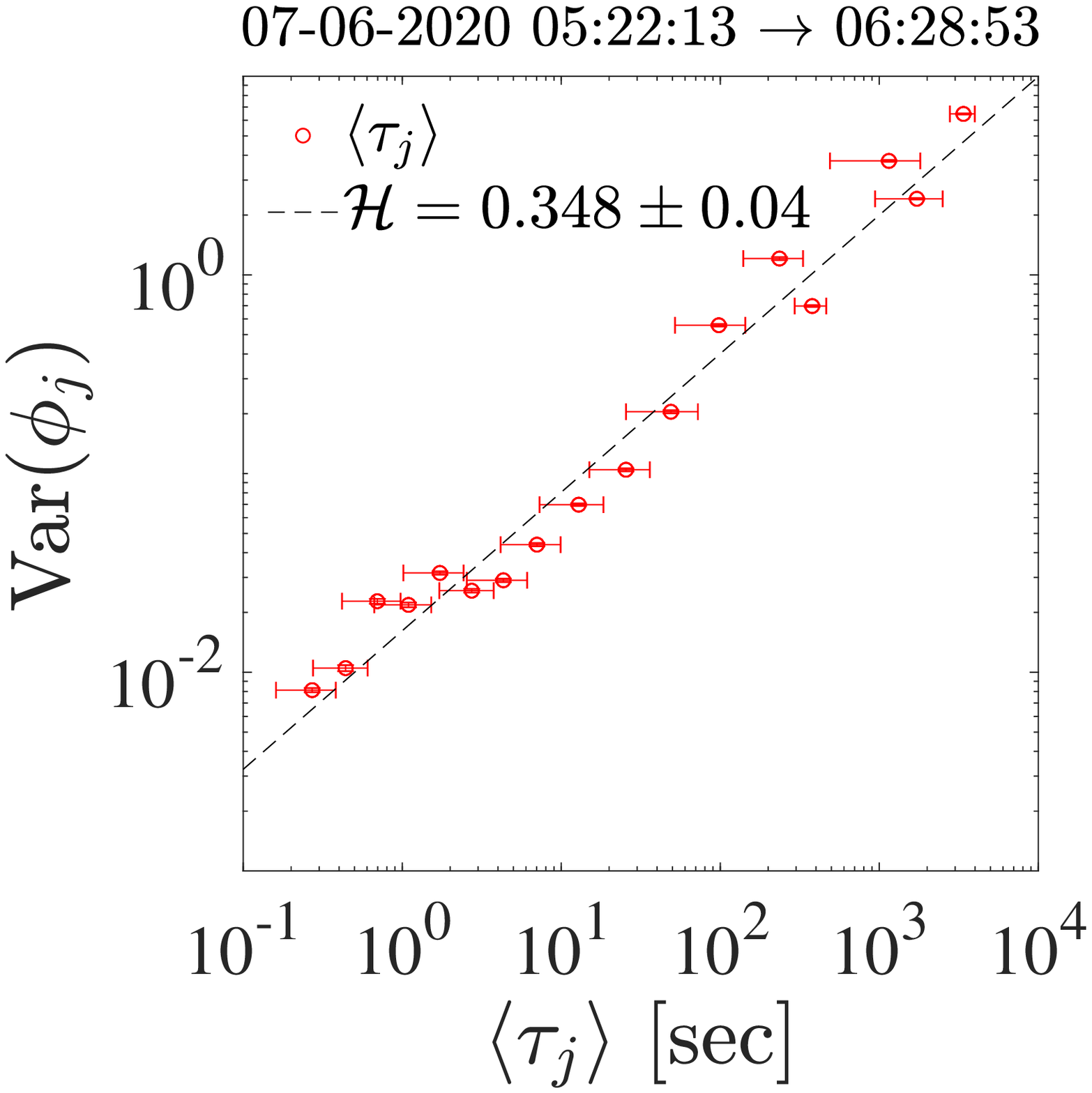}
		\caption{Left panel: an example of IMFs  $\phi_j(t)$ for sample $1$ (black lines). The bottom plot (red line) represents the residual $r_n(t)$. For better readability, only odd IMFs have been plotted.
		Central panel: average timescale $\langle\tau_j\rangle$ of each IMF of sample 1 as a function of the mode $j$. Error bars represent the $95\%$ confidence bounds. The dashed line is a least square fit obtained from the relation $\langle\tau_j\rangle = \alpha \gamma^j$.
		Right panel: IMF variance $\text{Var}(\phi_j)$ as a function of the average period $\langle\tau_j\rangle$ for sample $1$, with the associated $95\%$ confidence bounds. The dashed line represents the relation $\text{Var}(\phi_j)\sim \langle\tau_j\rangle^{2\mathcal{H}}$, being $\mathcal{H}$ the Hurst number (see the list for all intervals in Table~\ref{table2:H_slope_kurt}).
	}
		\label{fig1:emd_tau_H}
	\end{figure*}

	
Some examples of IMFs and the associated residual, as obtained trough the above EMD decomposition of the solar wind electron density $n_e$, are shown in the left panels of Figure~\ref{fig1:emd_tau_H} for sample $1$. 
For clarity, only odd IMFs have been plotted. 
Although the instantaneous frequency of the modes is variable, the figure highlights that each mode $j$ has a characteristic narrow frequency band, so that a mean period $\langle\tau_j\rangle$ and an associated variance $\text{Var}(\phi_j)$ can be properly defined.
Indeed, EMD acts intrinsically as a dyadic filter bank~\citep{Wu2004,Flandrin2004,Flandrin2004b,emd_book}, each IMF effectively capturing a narrow frequency band. 
However, the general features of the various IMFs depend on the specific process under analysis. For a turbulent field~\citep{Huang2008,Carbone16,Carbone2018,Carbone2020} or for a multifractal process~\citep{Carbone2010,Sorrisovalvo2017}, the characteristic mean period grows exponentially as $\langle\tau_j\rangle = \alpha \gamma^j$, where $\langle\cdot\rangle$ represents an ensemble average (in this case, average over time). The basis $\gamma$ can be evaluated empirically from the IMFs. For an exact dyadic decomposition $\gamma= 2$. 
Additionally, the variance of the IMF scales as a power of the mean timescale, $\text{Var}(\phi_j) \sim \langle\tau_j\rangle^{2\mathcal{H}}$. The scaling exponent $\mathcal{H}$ is the Hurst number, a parameter describing the persistence or anti-persistence of the fluctuations of the process under analysis~\citep{Nava2016,Carbone2019}.

The scaling of the mean period for sample $1$ is shown in the central panel of Figure~\ref{fig1:emd_tau_H}. The value $\gamma = 1.88 \pm 0.11$ obtained through a least-square exponential fit is compatible with the expected value for a dyadic decomposition, showing that EMD is correctly decomposing the data.
Moreover, the Hurst number $\mathcal{H}=0.348\pm0.04$ is obtained from the regression of the IMFs variance versus the average period, as shown in the right panel of Figure~\ref{fig1:emd_tau_H}. 
In this example, $\mathcal{H}$ is compatible with the standard value for classical fluid or magnetic turbulence  $\mathcal{H}=0.37$~\citep{benzi93,Arneodo1996,Bruno2016}, and indicates persistence of the fluctuations typical of intermittency.

Using the above procedure, the Hurst number has been evaluated for all $36$ intervals. The results, listed in Table~\ref{table2:H_slope_kurt}, show some variability that will be discussed and compared to other parameters in Section~\ref{sec:discussion}.

	\subsection{Arbitrary Order Hilbert Spectral analysis}
	\label{sec:spectra}
	
The Hilbert Spectral Analysis (HSA) is an extension of the basic EMD designed to characterize scale-invariant properties directly in the amplitude-frequency space~\citep{Huang2008}. It provides the equivalent of the power spectral density and high order moments of a field fluctuations (the structure functions), therefore representing a viable alternative to those standard tools.
After decomposing the field under study in its IMFs, the Hilbert transform of each mode is computed as:
	\begin{equation}
	\phi_{j}^\star(t) = \frac{1}{\pi}P\int_{-\infty}^{+\infty}\frac{\phi_{j}(t^\prime)}{t-t^\prime}
	\label{eq:hilbert_transf}
	\end{equation}
where $P$ is the Cauchy principle value. The Hilbert representation allows to extract a time-dependent instantaneous frequency $f_{j}(t) \equiv \tau_j^{-1}(t)$ and a time-dependent amplitude modulation $\mathcal{A}_{j}(t)$, by constructing the so called analytical signal $\Phi_j(t) = \phi_{j}(t)+i\phi_{j}^\star(t) \equiv \mathcal{A}_{j}(t)e^{i\theta_j(t)}$~\citep{Cohen95}.
Here $\mathcal{A}_{j}(t) = |\Phi_j(t)| \equiv \sqrt{\phi_{j}^2(t)+{\phi_{j}^\star}^2(t)}$ and $\theta_j(t) = \arctan[\phi_j^\star(t) / \phi_j(t)]$ are the instantaneous amplitude modulation and the instantaneous phase oscillation, respectively (the instantaneous frequency being defined as $f_j(t) = 2\pi^{-1}d\theta_j(t)/dt$)~\citep{long1995,Cohen95,Flandrin_book}.
	
After rewriting the original signal in terms of $\mathcal{A}_j$ and $\theta_j$, $n_e = \text{Real}\left[ \sum_j\mathcal{A}_j(t)\exp\left(i\int f_i(t)dt\right) \right]$, the energy as a function of the instantaneous frequency $f$ and time, can be defined as $h(f) = \int_0^\infty H(f,t) dt$, or the marginal integration of the Hilbert spectrum $H(f,t) \equiv \mathcal{A}^2(f,t)$ (being $H(f,t)$ a representation of the original signal at the local level) ~\citep{Huang1998,Huang2009b}.
An equivalent definition of $H(f,t)$ can be obtained from the joint probability density function of the instantaneous frequency and amplitude $P(f_j, \mathcal{A}_j)$~\citep{long1995}, extracted from the IMFs.
In this case, the Hilbert marginal spectrum is the second statistical moment of such distribution, analog to the Fourier spectral energy density:
	\begin{equation}
	h(f)\equiv\mathcal{L}_2(f) = \int_0^\infty P(f,\mathcal{A})\mathcal{A}^2d\mathcal{A} \ .
	\label{eq:L2}
	\end{equation}
The above definition can be then generalized to any arbitrary moment $q\geq0$, representing the analogous of the standard structure functions of the fluctuations:
	\begin{equation}
	\mathcal{L}_q(f) = \int_0^\infty P(f,\mathcal{A})\mathcal{A}^qd\mathcal{A} \ .
	\label{eq:Lq}
	\end{equation}	
Equations~(\ref{eq:L2}-\ref{eq:Lq}) are used here to estimate the spectral and intermittency properties of the electron density turbulence of the $36$ sub-intervals of this study. 

\begin{figure*}
		\centering\includegraphics[scale=0.33]{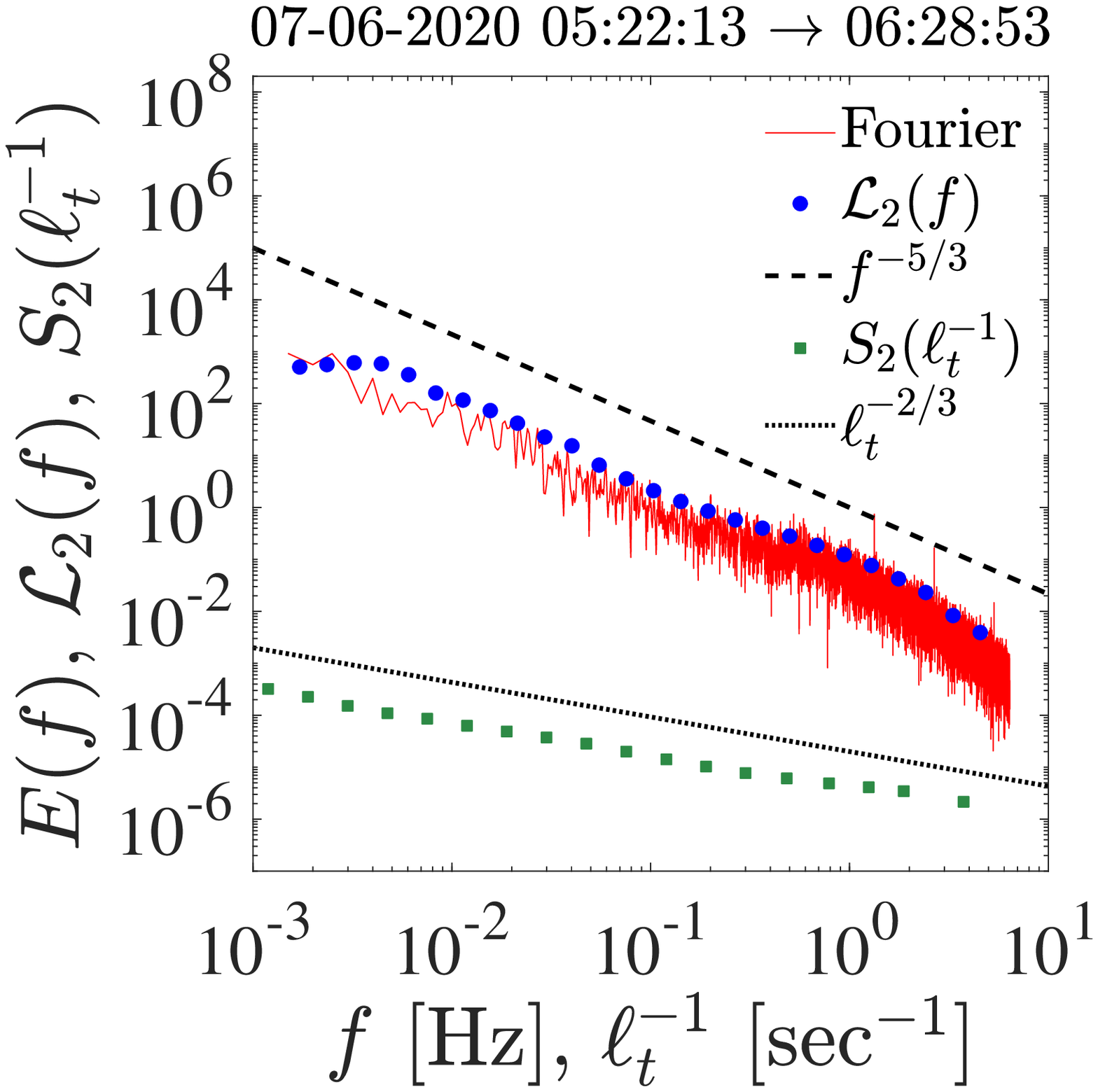}\includegraphics[scale=0.33]{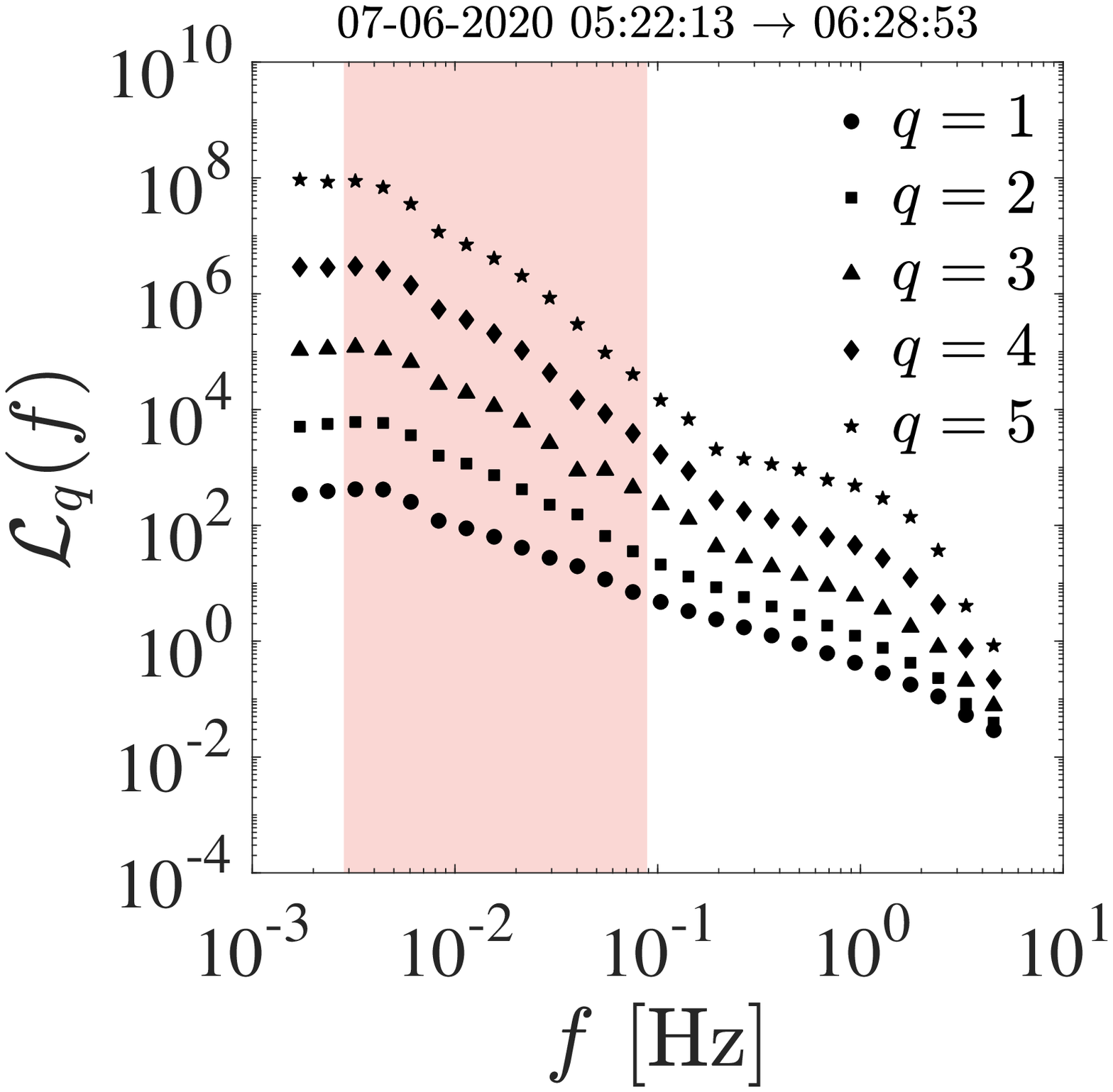}\includegraphics[scale=0.33]{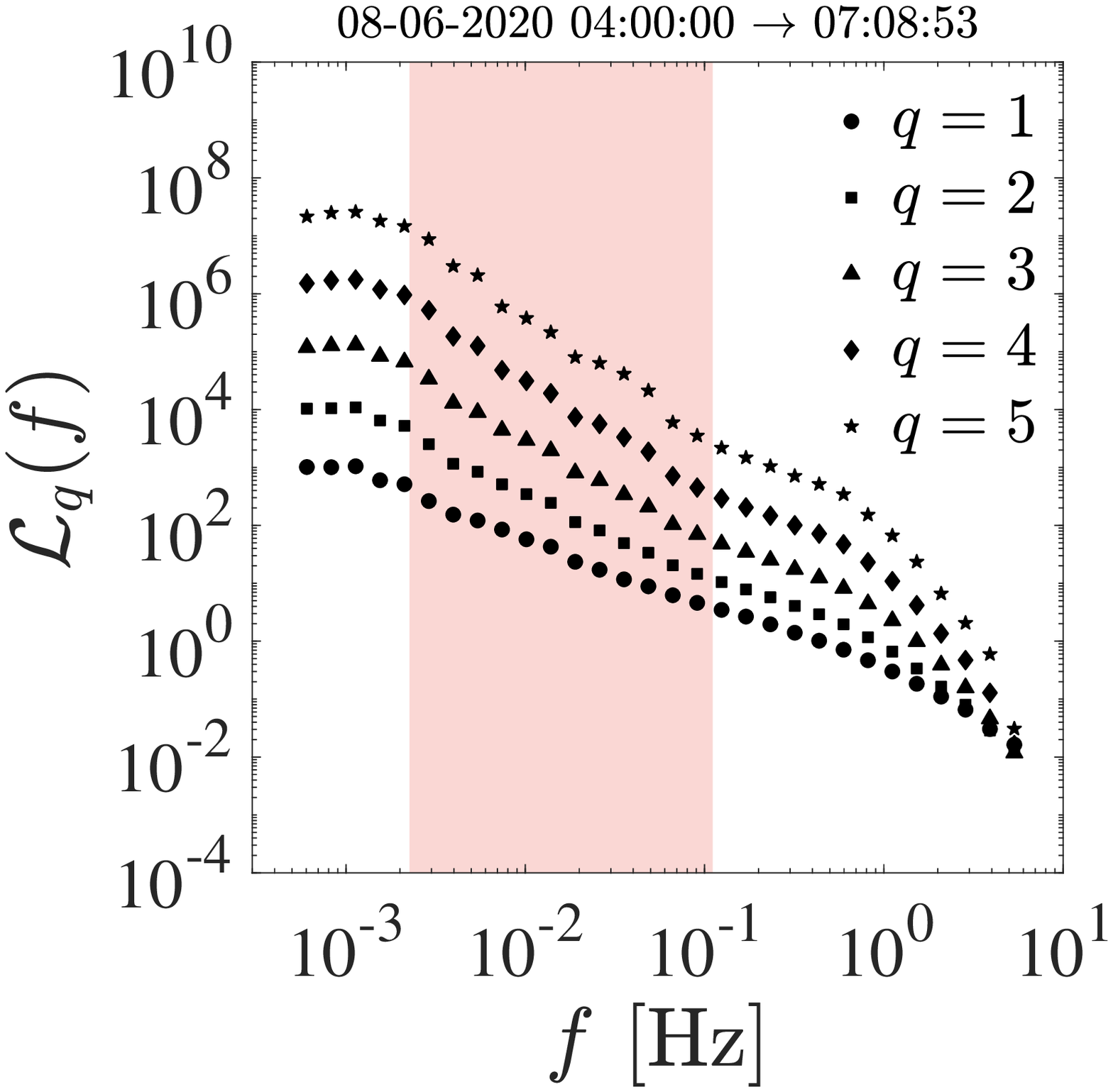}
		\caption{Left panel: second order Hilbert spectrum $\mathcal{L}_2(f)$ (blue circles), the classical Fourier PSD~\citep[][]{Welch1967} $E(f)$ (red line), and the second-order structure function $S_2(\ell_t)$
		(green squares, plotted as a function of the inverse timescale $\ell_t$), for sample 1. 
		Power-scaling is present in the same frequency range for all methods. The dashed line represents the classical Kolmogorov scaling $\mathcal{L}_2(f) \sim f^{-5/3}$, while the dotted line
		represent the expected scaling for the second-order structure function $S_2(\ell_t)\sim\ell_t^{\zeta(2)}$ ($\zeta(2) = \beta_2 - 1$).
		Central panel: generalized Hilbert spectra $\mathcal{L}_q(f)$ for $q\in[1,5]$, obtained for sample 1. The curves have been vertically shifted for clarity. The shaded area represents the frequency range of the bootstrapping least-square fit.
		Right panel: same as in the central panel, for sample 9. The power behavior is still present, but the power-law exponents are considerably different.}
		\label{fig2:spettri}
	\end{figure*}

The left panel of Figure~\ref{fig2:spettri} shows one example of the equivalent spectrum $\mathcal{L}_2(f)$ (blue circles), obtained through the HSA described above, for sample $1$. 
For comparison, the classical power spectral density $E(f)$ (red line), evaluated through the Welch's method~\citep{Welch1967}, and the second-order structure function $S_2(\ell_t^{-1})$ are also shown. Both spectra clearly display power-law scaling $E(f),\mathcal{L}_2(f) \sim f^{-\beta_2}$, in the frequency range $f\in[3\times10^{-3},10^{-1}]$, compatible with the typical inertial range of time-scales. Similarly, the second-order structure function also show a very clear power law scaling. In all three cases, the power-law scaling of $\mathcal{L}_2(f)$ is compatible with the standard Kolmogorov predictions ($S_2(\ell_t)\sim\ell_t^{2/3} \to E(f)\sim f^{-5/3}$), also shown in the left panel of the figure.

\begin{figure}
		\centering
		\includegraphics[scale=0.4]{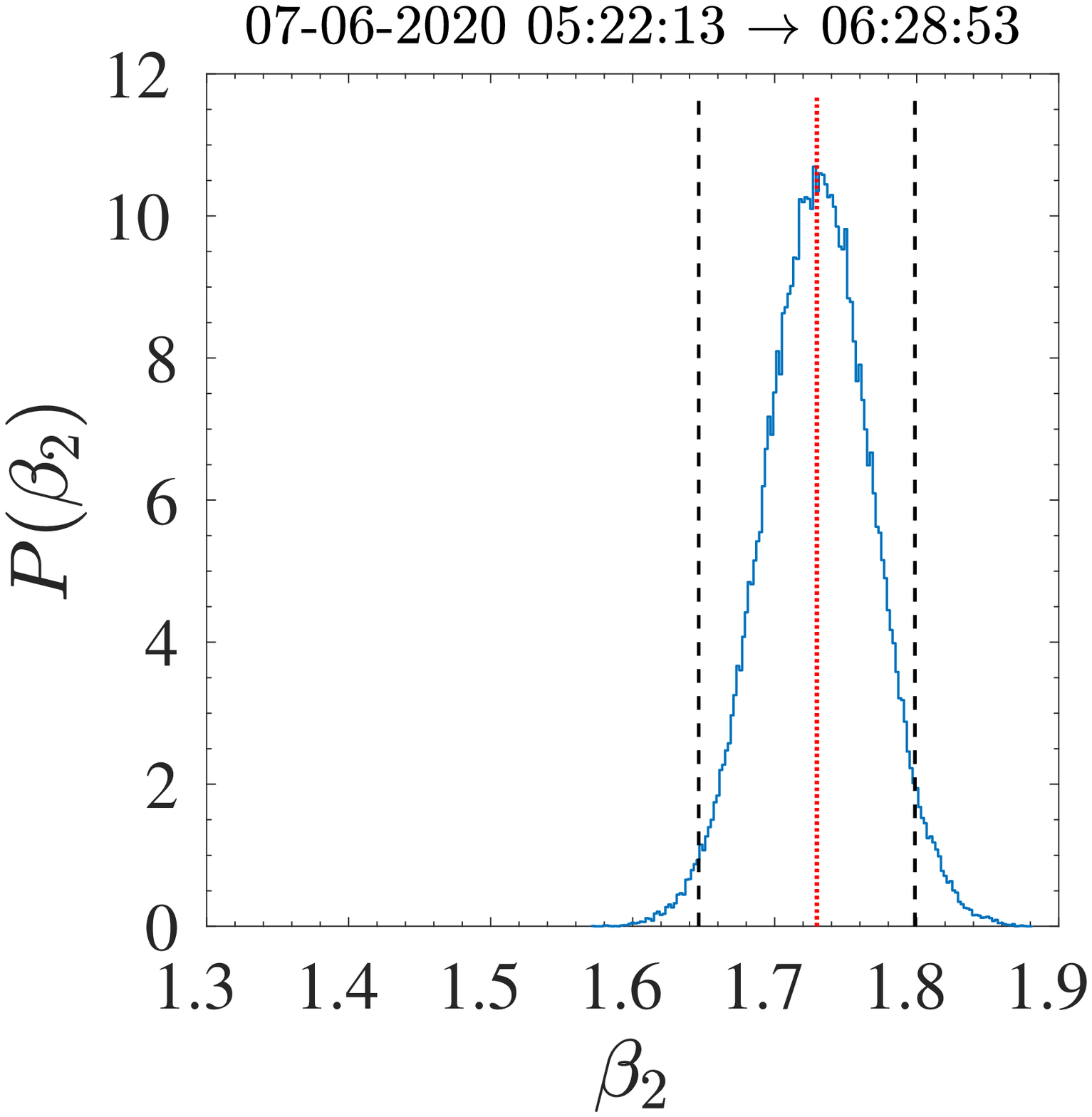}
		\includegraphics[scale=0.4]{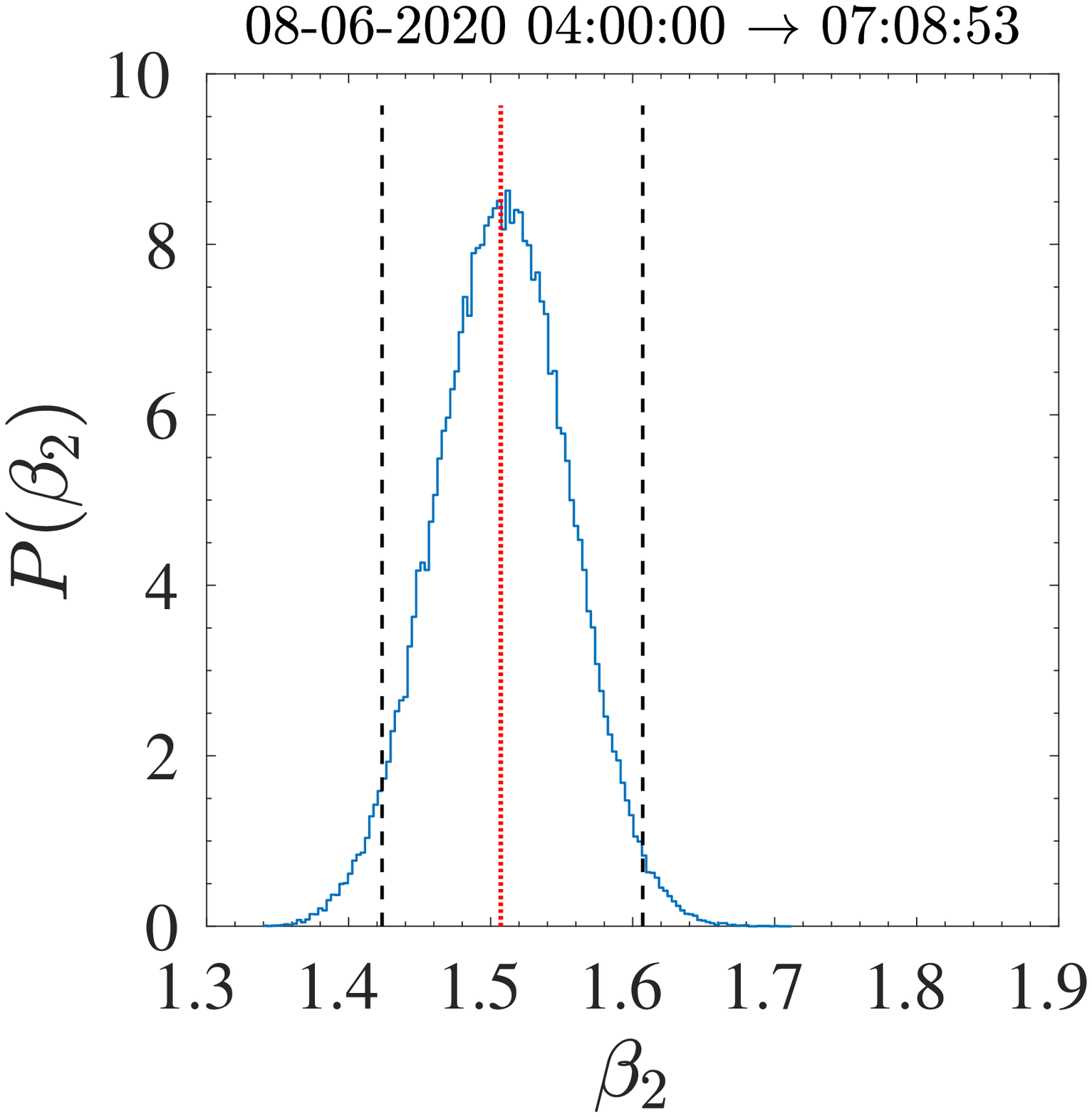}
		\caption{Probability distribution function $P(\beta_2)$ of the scaling exponents $\beta_2$, constructed	via bootstrap resampling, for sample 1 (upper panel), and sample 9 (lower panel).
		In both panels, the vertical dotted line represents the median of the distribution ($50$th percentile), while the vertical dashed bars indicate the $95\%$ confidence interval (enclosed between $2.5$th and $97.5$th percentile).}
		\label{fig5:dist_beta2}
	\end{figure}

The HSA equivalent spectrum $\mathcal{L}_2(f)$ displays a slightly better power-law scaling than the traditional Fourier spectral density, which has some weak amplitude modulation. Thanks to the local nature of the EMD analysis, the sources of such modulation can be removed, allowing to obtain a more precise determination of the spectral scaling exponents. This corresponds to isolating the properties of the turbulent cascade from the possible effects of the instrumental noise, and of the larger-scale energy inhomogeneity~\citep{Huang2010,Carbone2018,Telloni2019}.
Note that the power-law scaling range can vary for the various samples, but always includes at least one decade of scales. The two spectral estimators are not always nicely superposing, with the Fourier spectrum occasionally presenting stronger modulations (not shown). This suggests the possible presence of large-scale modulations, which may affect the Fourier spectra but are well controlled by the HSA.
In order to achieve robust estimate, the scaling exponent of the $q$-th Hilbert spectrum $\mathcal{L}_q(f)$ was 
evaluated via the residual resampling (bootstrapping) procedure~\citep{Bootstrap_book,Carbone2020b}. First, a least square fit is performed on each $\mathcal{L}_q(f)$, then the residuals are randomly resampled and added to the fit, generating a new data-set (replica). 
The replica is then fitted again, and the procedure is repeated for a number of times, in this case $N_{boot} = 10^4$~\citep{Boos2010,Wilcox2010}. Such large number of replica is necessary for correctly evaluating confidence or prediction interval, wile in general simple statistical tests require a smaller number ($N_{boot} \in [50, 100 ]$)~\citep{Dale2012}.
The probability distribution of the exponents $\beta_q$ (or the scaling exponents $\beta_q - 1$) obtained from the $N_{boot}$ least square fits is finally used to estimate the $50$-th percentile (median of the distribution), used as the best estimate of the exponents, and the statistical error ($95\%$ confidence interval)~\citep{Efron2015,Wilcox2010}.
Figure~\ref{fig5:dist_beta2} shows two examples of the distribution $P(\beta_2)$ estimated for samples $1$ and $9$ using the bootstrapping technique. 
The red dotted vertical line represents the median, and the black dashed lines represents the statistical error. 
In the examples shown in the figure, the median of the distribution provides the scaling exponent $\beta_2 = 1.72\pm0.14$ for sample $1$ (the error indicating the $95\%$ confidence interval), in excellent agreement with the classical scaling for the fully developed hydrodynamic turbulence $\beta_2 \approx 1.7$~\citep[][]{benzi93}, and a shallower $\beta_2 = 1.51\pm0.10$ for sample $9$, consistent with the Iroshnikov-Kraichnan spectrum for Alfv\'enic turbulence~\citep{iroshnikov1964,kraichnan1965}.
Values of $\beta_2$ were obtained for all $36$ samples. A discussion about these values, collected in Table~\ref{table2:H_slope_kurt}, will be provided in next Subsection and in Section~\ref{sec:discussion}.


	\subsection{Scaling exponents and intermittency analysis}	
    \label{sec:groups}

In the framework of the standard Kolmogorov turbulence, 
a direct link exists between the Fourier spectral exponent $E\sim f^{-\beta}$ and the scaling exponent of the second-order structure function, $S_2 \sim \ell_t^{\zeta(2)}$, so that $\beta-1 = \zeta(2)$.
This relationship can be extended to any moment order $q$ of the generalized Hilbert spectra $\mathcal{L}_q\sim f^{-\beta_q}$, yielding the generalized scaling exponents $\xi(q) \equiv \beta_q - 1$
~\citep{Huang2010,Carbone16}.
These are the analogous of the scaling exponents $\zeta(q)$ obtained using the standard structure functions~\citep{Frish1995,benzi93,Arneodo1996}, and can be used to retrieve quantitative information on the properties of turbulence. 
Additionally, the structure function scaling exponents are linked to the Hurst number via the relation $\zeta(q) = q\mathcal{H}$ (in absence of intermittency corrections). This allows an alternative estimate of the Hurst number using, for example, the first-order exponent $\mathcal{H}=\beta_1-1\equiv \xi(1)$. 

The central and right panels of Figure~\ref{fig2:spettri} show two examples of $\mathcal{L}_q(f)$ (for intervals 1 and 9), obtained from Equation~\ref{eq:Lq} up to the $5$th order, for the electron density $n_e$ in samples $1$ and $9$.  All curves presents good power-law scaling for all orders, approximately in the inertial range of frequencies (shaded areas). 
The associated generalized Hilbert spectra scaling exponents $\beta_q$, and hence the equivalent structure-function scaling exponents $\xi(q)$, were obtained through the bootstrapping procedure described above and were used to determine the intermittency properties of the electron density. 
In order to check the quality of the procedure, the first-order exponents were initially used to obtain the alternative estimate of the Hurst number. For the example of sample 1, the value $\mathcal{H}=0.32\pm0.06$ was obtained, in good agreement with the value obtained through the regression of the IMF variance versus the average timescale, illustrated in Figure~\ref{fig1:emd_tau_H}. This was consistently observed for all the 36 samples.
The power-law exponents of $\mathcal{L}_q(f)$ are visibly different in the two intervals shown in the central and right paneThe power-law exponents of $\mathcal{L}_q(f)$ are visibly different in the two intervals shown in the central and right panels of Figure~\ref{fig2:spettri}. 

ls of Figure~\ref{fig2:spettri}. 
For $q=2$, this can be also more quantitatively noticed by comparing the distributions and median values shown in Figure~\ref{fig5:dist_beta2} for the same intervals. 

\begin{figure*}
		\includegraphics[scale=0.35]{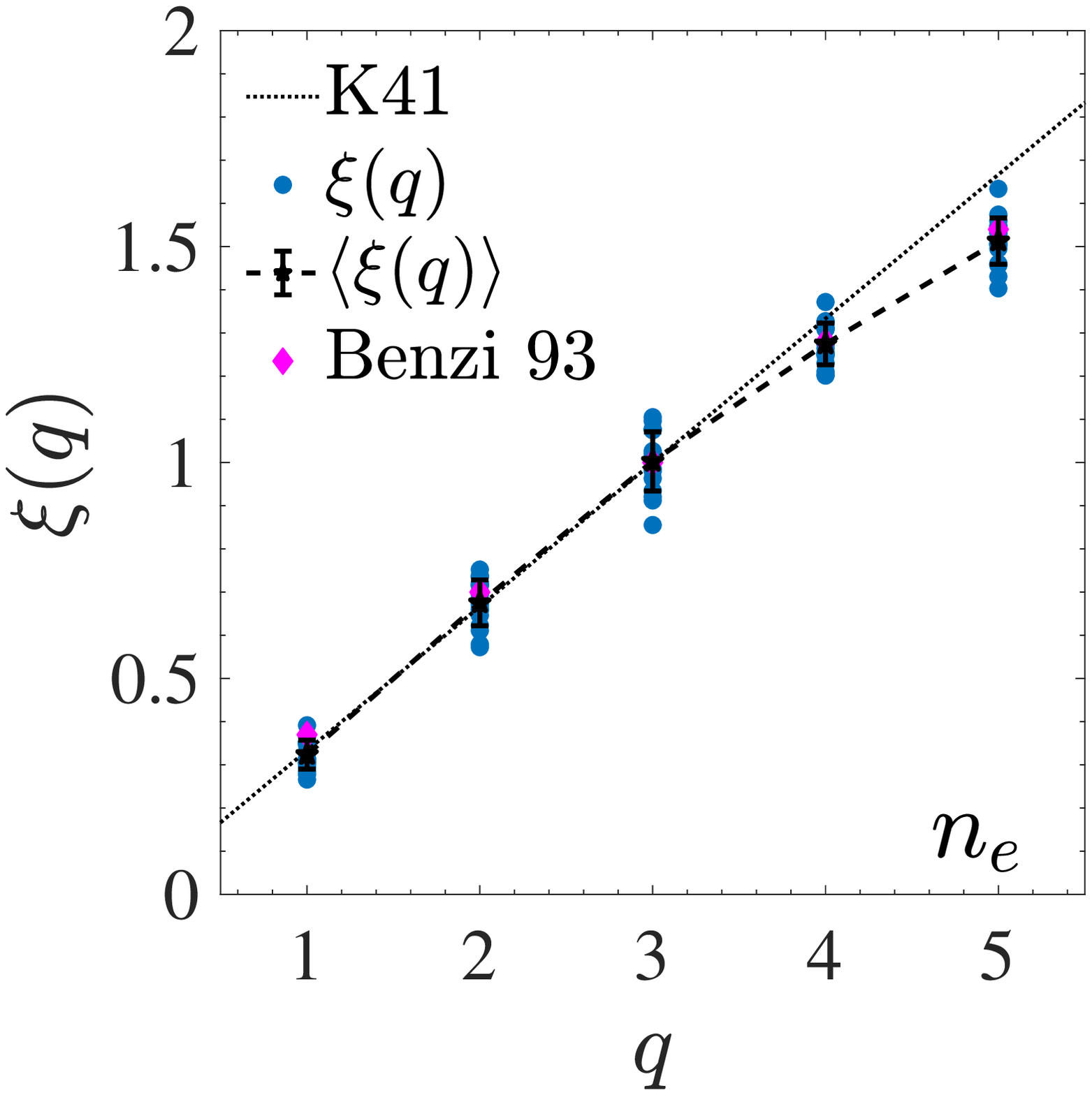}\includegraphics[scale=0.35]{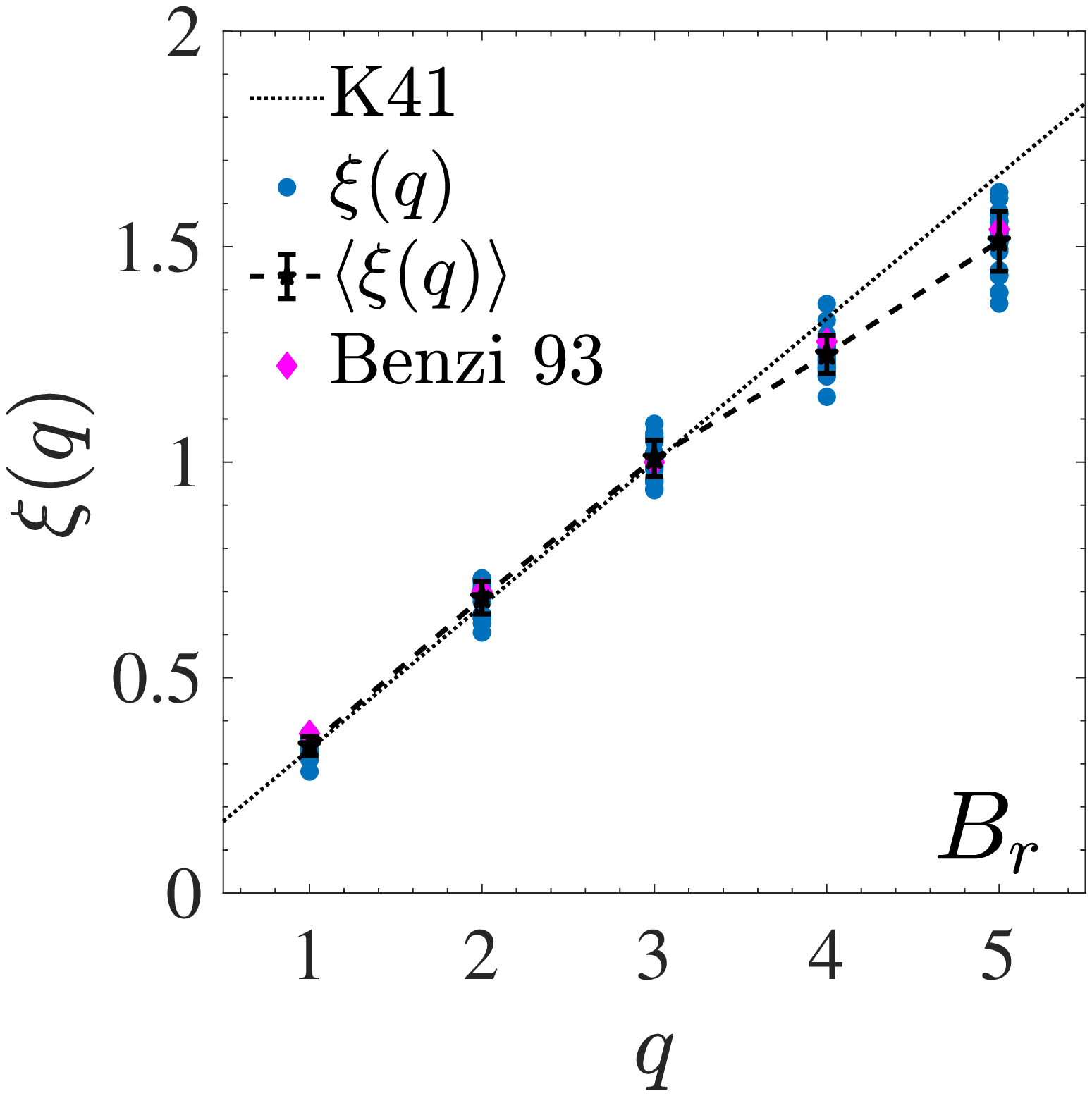}\includegraphics[scale=0.35]{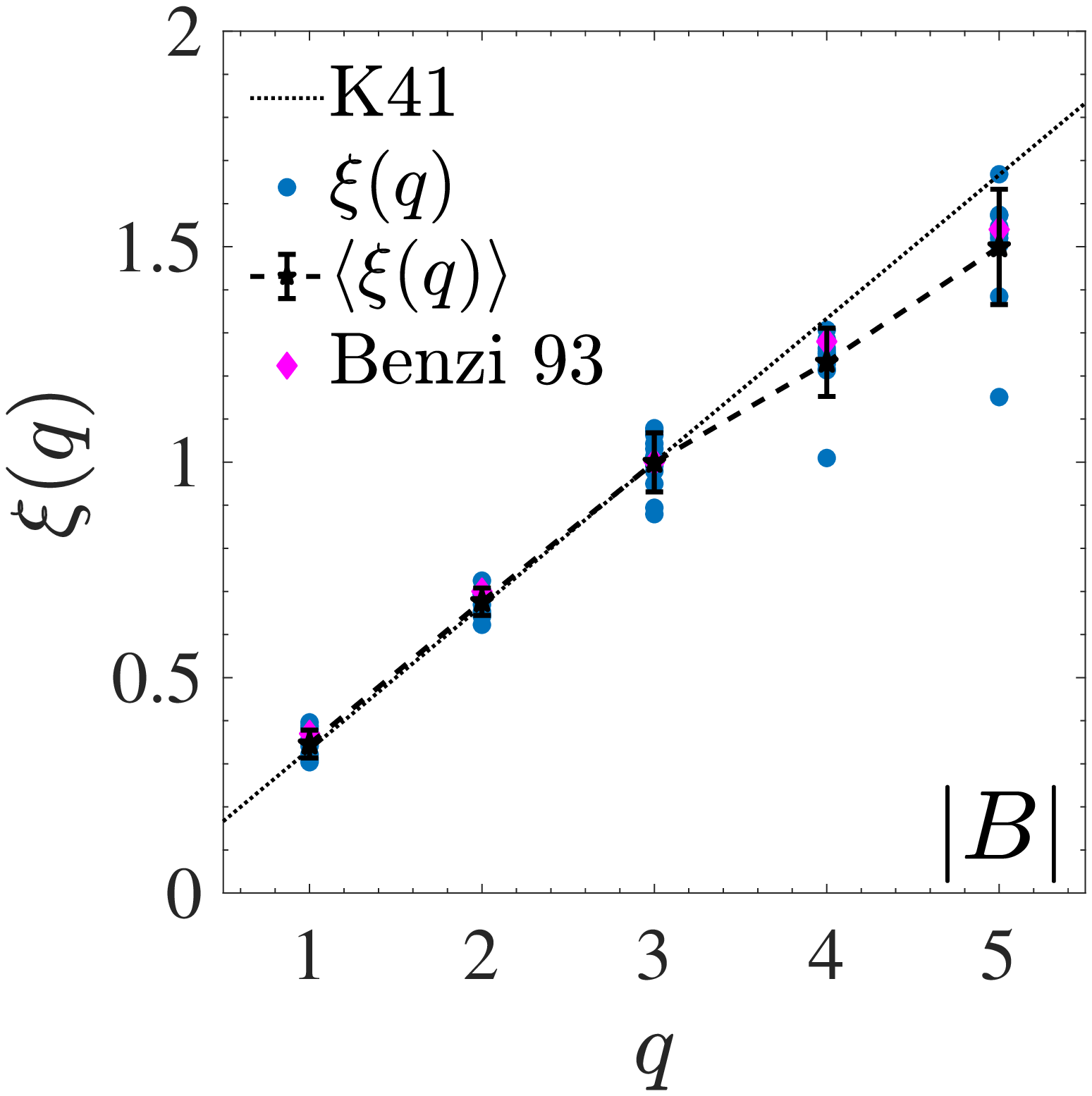}
		\includegraphics[scale=0.35]{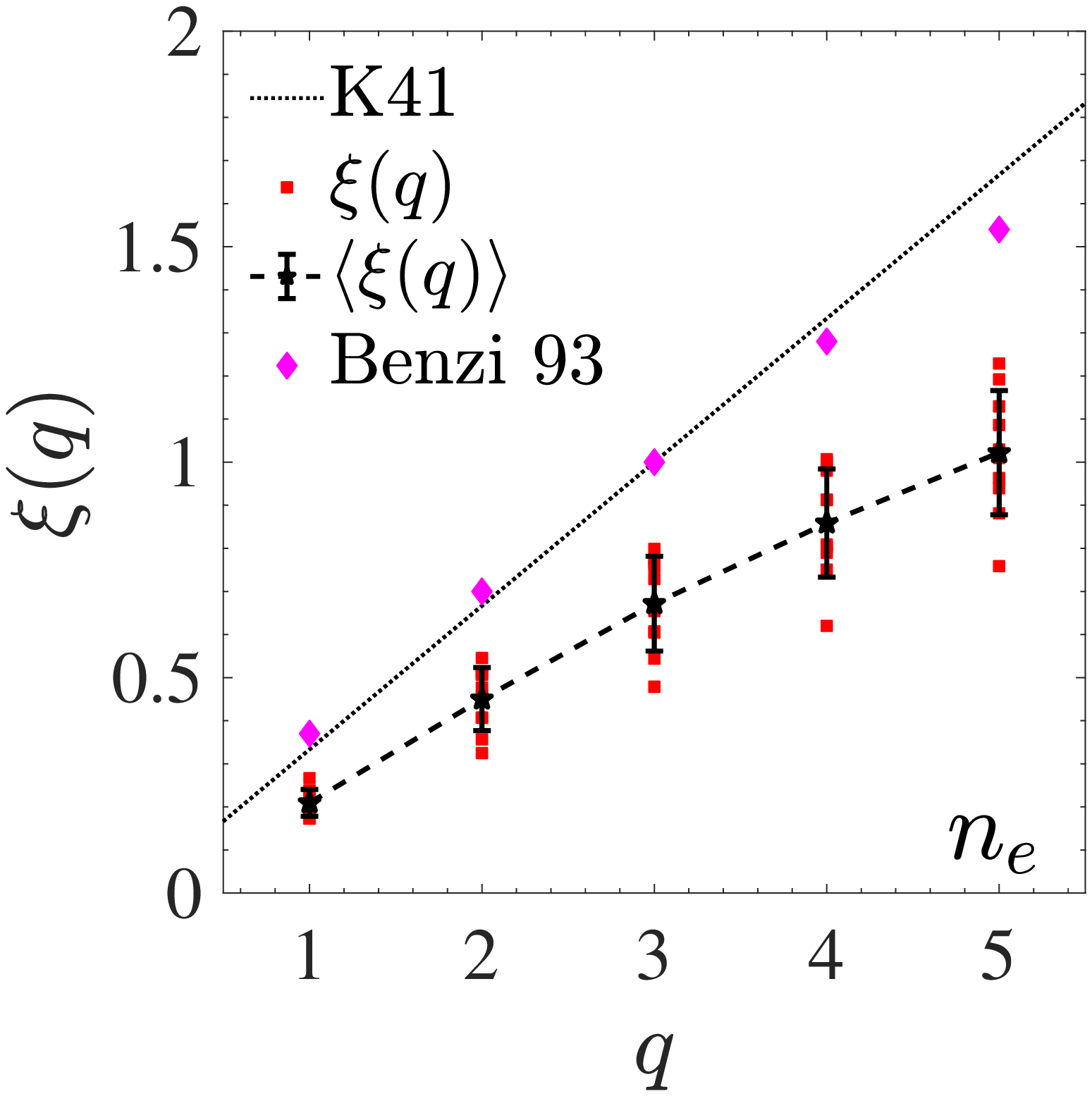}\includegraphics[scale=0.35]{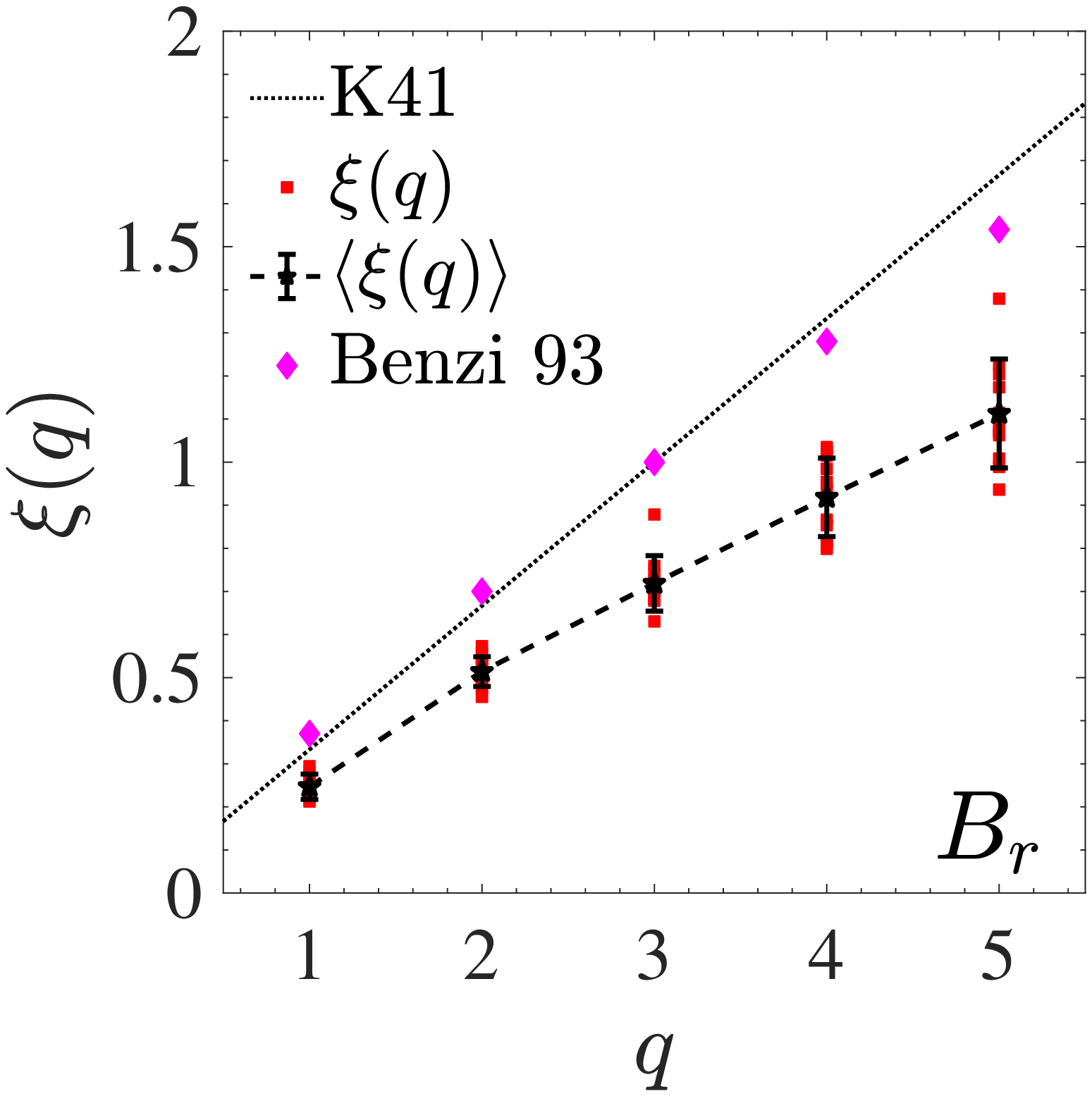}\includegraphics[scale=0.35]{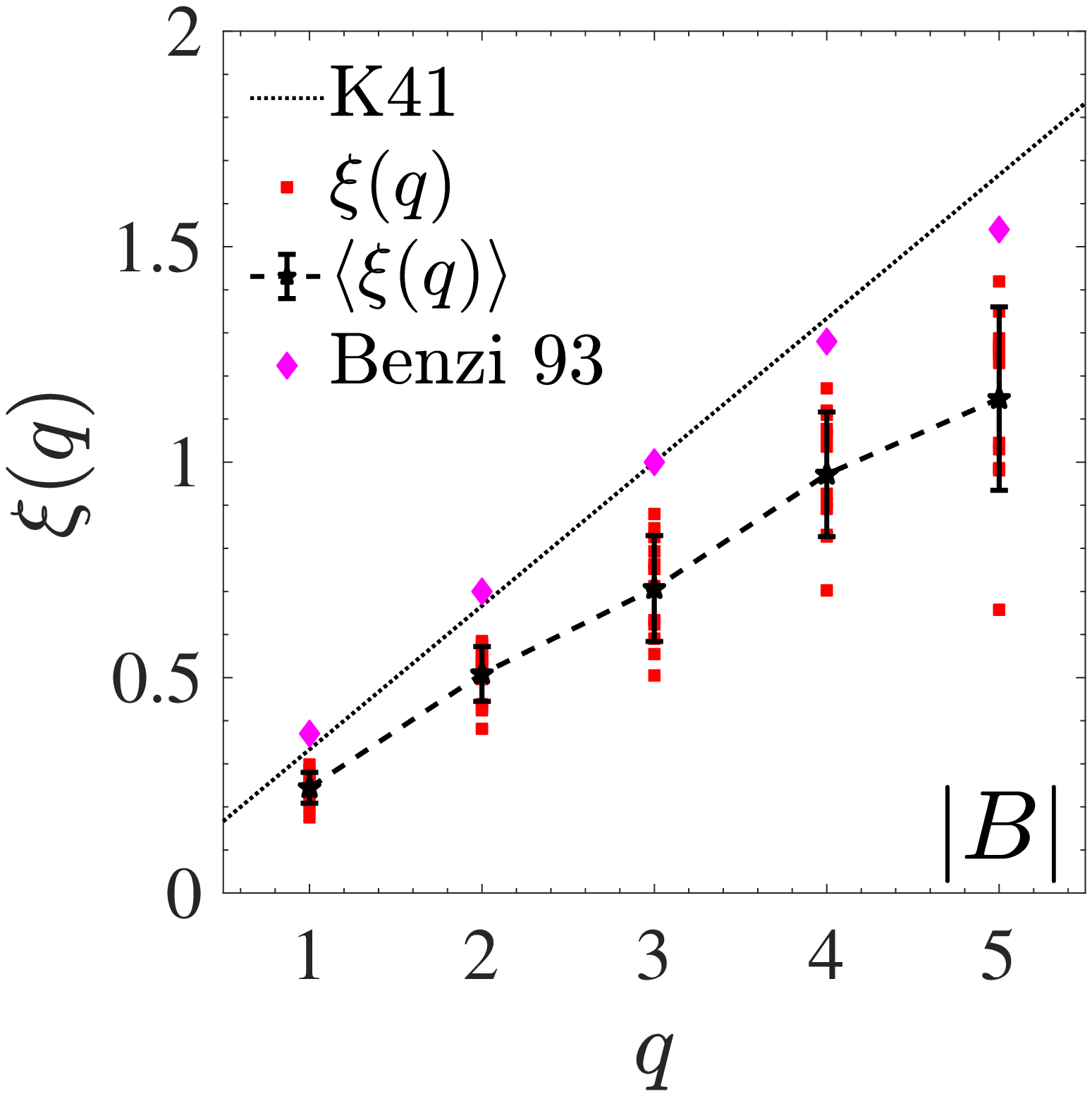}
		\caption{Top-left panel: HSA equivalent scaling exponents $\xi(q)$ for group 1 (blue circles) and their average (black crosses). The exponents $\zeta(q)$ for velocity fluctuations in the inertial range of hydrodynamic turbulence (red squares) are shown for reference~\citep{benzi93}. The dashed line represents the theoretical expectation $\zeta(q) = q/3$~\citep{K41}.
		Top-center and top-right panels: equivalent scaling exponents for the magnetic field radial component $B_r$ and magnitude $|B|$, for group 1. 
		Bottom panels: same as the top panels, for the intervals of group 2.}
		\label{fig4:exps}
	\end{figure*}

The scaling exponents $\xi(q)$ are shown in the top- and bottom-left panels of Figure~\ref{fig4:exps} for most of the samples, separated in two groups as will be described in the following. The scaling exponents for the magnetic field magnitude $|\mathbf{B}|$ (central panels) and radial component $B_r$ (right panels) are shown for comparison.  
For all cases included in the figure, the curvature of the exponents with respect to the linear prediction of the non-intermittent Kolmogorov phenomenology is evident. This universal behaviour of turbulence is due to the effects of the intermittency, or anomalous dissipation~\citep{kolmogorov1962,Schmitt1994,Schmitt2003,Bruno2016,Carbone2019}, and are related to the multifractal nature of the turbulent cascade~\citep{Meneveau1991,Davis1994,Sorrisovalvo2017}.	
In all panels, the scaling exponents from a classical measure of fluid intermittent turbulence are also shown for comparison~\citet{benzi93}. 
In order to describe their intermittency behaviour, the 36 samples were then separated in three groups, according to the behaviour of the generalized scaling exponents and of the Hurst number, with respect to the standard fluid turbulence reference. Note that, for each sample, the same behaviour is consistently observed for density and magnetic field.

The first group (group 1, including 21 intervals) displays the typical statistical features of fully developed turbulence (top panels of Figure~\ref{fig4:exps}). 
For these samples, the scaling exponents are consistent with the reference values from fluid turbulence, and are well described (not shown) by models of intermittent turbulence~\citep[e.g. the p-model by][not shown]{Meneveau1991}. 
In particular, the equivalent spectral exponent $\beta_2 > 1.55$ is always compatible with the Kolmogorov scaling ($\beta_2 = 5/3$). 
Furthermore, for samples in this group $\mathcal{H}\in[0.30,0.39]$, compatible with the classical value obtained for ordinary fluid turbulence $\mathcal{H}=0.37$. 
From these observations, we conclude that the samples in group 1 are characterized by a standard turbulence, with the expected power-law spectra, presence of small-scale intermittent structures and anti-persistent fluctuations. 

In the second group (group 2, including 11 samples), the scaling exponents are characterized by more extreme deviation from the linear prediction, and by much smaller values (bottom panels of Figure~\ref{fig4:exps}). These exponents also deviate considerably from the fluid reference, and their order dependence cannot be described by standard models of turbulence (not shown). The equivalent spectral exponents $\beta_2 < 1.55$ are shallower than the Kolmogorov scaling, and in some cases are compatible with the Iroshnikov-Kraichnan scaling. 
The Hurst number is also consistently smaller than for standard turbulence, $\mathcal{H}\in[0.16,0.26]$. 
These observations suggest that, unlike in group 1, the density and magnetic fluctuations in these samples may not be generated by a standard turbulent cascade. Some other processes might coexist, modifying the statistics. Note that the EMD-based analysis constrains the effects of
finite-size sample, poor stationarity and large-scale structures effects. Therefore, it can be claimed that the observed features might be related to the presence of small- or inertial-scale fluctuations that are not generated only by a turbulent cascade. 

The third group (group 3, not shown) includes 4 samples that do not show clear power-law scaling of the generalized Hilbert spectra for all orders, so that not all the scaling exponents are available. These intervals, associated with small Hurst number and spectral exponent, are therefore not representative of turbulence. 

\begin{figure}
		\centering\includegraphics[scale=0.48]{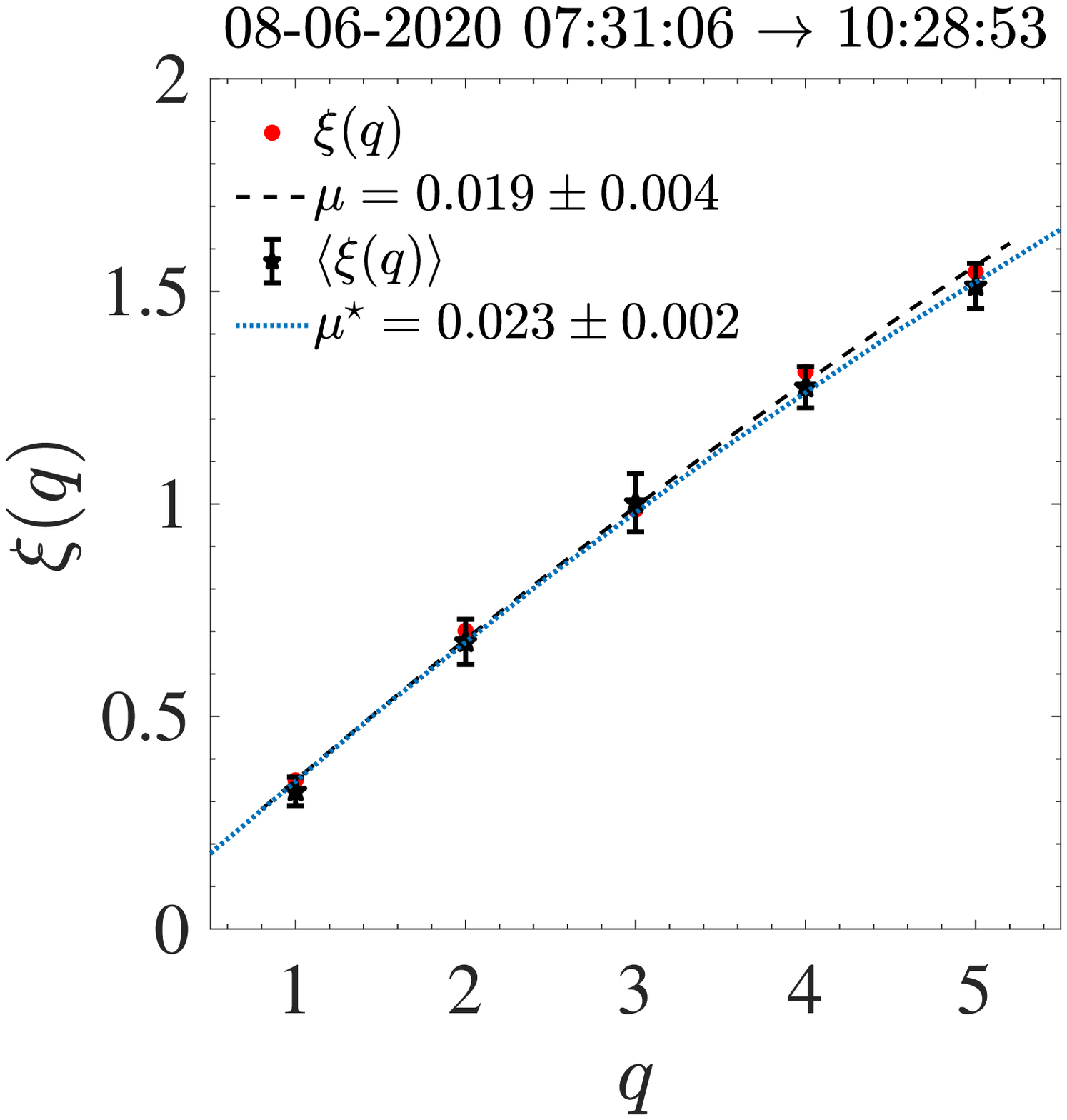}
		\caption{Least square fit of the scaling exponents $\xi(q)$ via the log-normal cascade model~(\ref{eq:intermit}) (lines), for sample 10 (red circles) and for the average of all exponents (black stars). The fitting parameter $\mu$ is in agreement with that of classical hydrodynamic turbulence $\mu \approx 0.02$.}
		\label{fig5:mu_fit}
	\end{figure}

The multifractal nature of the fluctuations can be quantitatively described fitting the scaling exponents $\xi(q)$ to a log-normal model~\citep{Schmitt2003,Medina2015}: 
    \begin{equation}
	\xi(q) = q\mathcal{H} - \frac{\mu}{2}\left(q^2 - q\right) \ .
	\label{eq:intermit}
	\end{equation}
The model is able to describe standard intermittent turbulence when the curvature parameter $\mu \approx 0.02$. For other multifractal processes, not generated by a nonlinear turbulent cascade, different values can be obtained.  
One example of log-normal model fit of the equivalent scaling exponents $\xi(q)$ is shown in Figure~\ref{fig5:mu_fit} for group 1 (sample 10, red circles), giving  $\mu = 0.019\pm0.004$. 
The model was also fitted to the exponents obtained ensemble-averaging all samples of group 1 (stars), providing $\mu^\star = 0.023\pm0.002$. 
Alternatively, the average parameter computed using the results of the fit of all samples of group 1 is $\langle \mu\rangle_1 = 0.028\pm0.010$ (the error representing the standard deviation).
All the above values are in good agreement with those observed for standard turbulence. Similar values were obtained for the magnetic field magnitude $\left|B\right|$ (e.g., $\langle \mu\rangle_1 = 0.027 \pm 0.010$).
The exponents for group 2 were also fitted to relation~\ref{eq:intermit}. In that case, the resulting parameters were generally smaller, with average $\langle \mu\rangle_2 = 0.013\pm0.01$. 
The parameters from all 36 intervals are plotted in one of the panels of Figure~\ref{fig:scatter} and will be discussed in Section~\ref{sec:discussion}. 
While there is a considerable spread in both groups, the parameters for group 2 appear generally smaller, confirming that the fluctuations have peculiar, strongly multifractal structures that do not simply originate from a nonlinear turbulent cascade.

\begin{figure}
		\centering\includegraphics[scale=0.5]{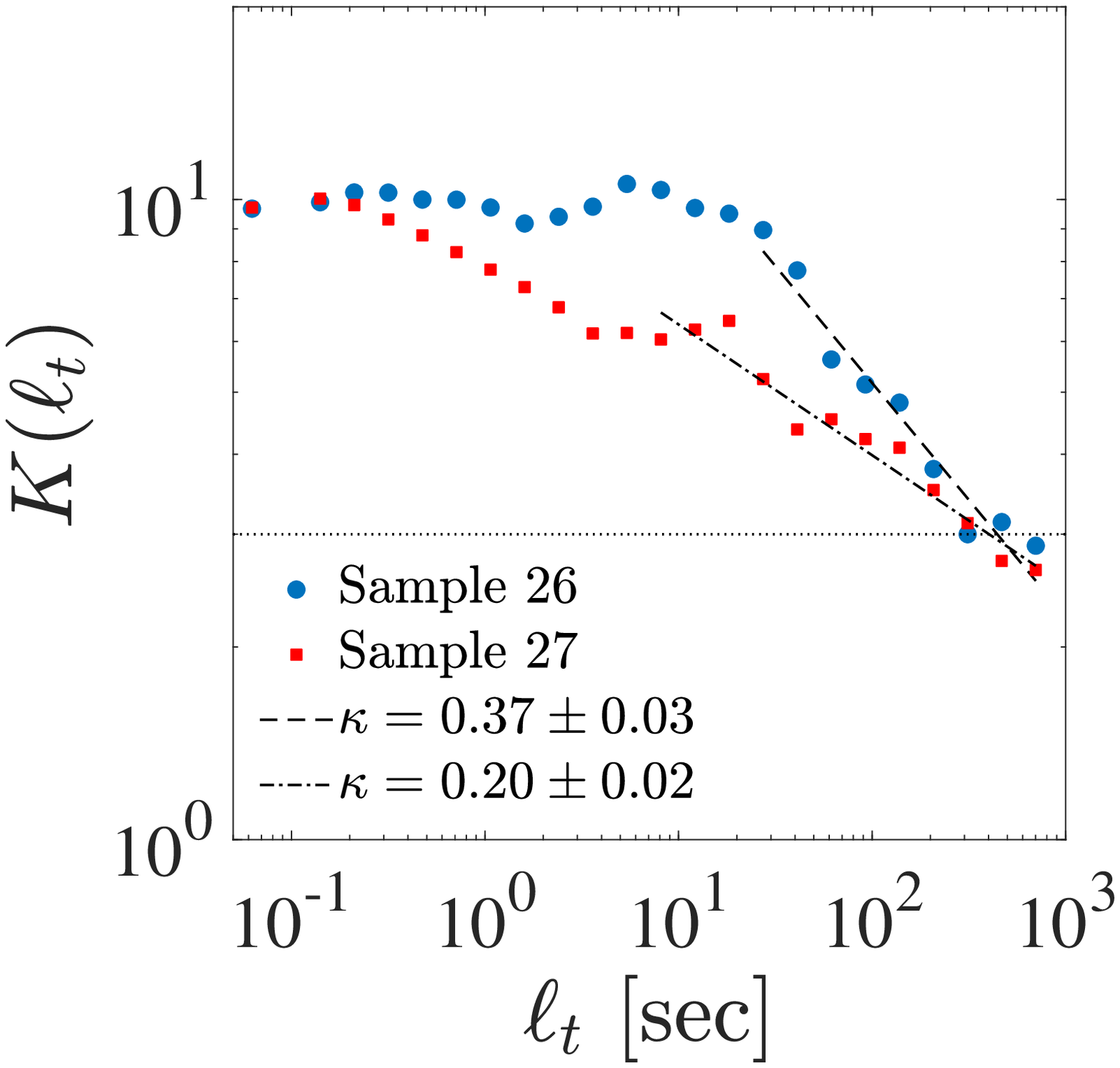}
		\caption{Scale-dependent kurtosis $K(\ell_t)$ for samples 26 (group 1) and 27 (group 2). Power-law fits $K(\ell_t)\sim \ell_t^{-\kappa}$ give the indicated exponents $\kappa$. The Gaussian value $K=3$ is also shown (horizontal dotted line).}
		\label{fig:k}
	\end{figure}

Finally, in order to include also one example of standard data analysis technique for intermittent turbulence, we have estimated the kurtosis of the fluctuation distribution, namely the ratio between the fourth-order and the squared second-order structure functions $K(\ell_t)=S_4 / S_2^2 \sim \ell_t^{-\kappa}$~\citep{Frish1995,DudokDeWit2013}. 
The kurtosis provides information on the shape of the distribution of the scale-dependent fluctuations. At large scales (comparable with the system correlation scale) the Gaussian value $K=3$ is typically observed. As the scale decreases, the inhomogeneous turbulent cascade generates intermittent structures, so that the distribution deviates from Gaussian, corresponding to increasing $K$. The scaling properties of turbulence result in the power-law scaling of $K$ in the inertial range. The scaling exponent $\kappa$ is a good measure of the efficiency of the cascade, namely how rapidly the small-scale structures are generated. This depends on the nature of the nonlinear interactions and can be used as a quantitative measure of intermittency~\citep{Castaing1990,Carbone2014}. 
The kurtosis was estimated for all intervals, and a power-law fit was performed whenever a long enough scaling range was observed.
An example of $K$ with power-law fit is shown in Figure~\ref{fig:k}. The resulting fitting  parameters $\kappa$ are collected in Table~\ref{table2:H_slope_kurt}. 
)

	\section{Observation of ion cyclotron waves}
	\label{sec:waves}

Observations of solar wind data often reveals the presence of wave activity near the end of the MHD inertial cascade range, and close to the kinetic plasma range.
These are typically identified as kinetic Alfv\'en waves (KAW), or ion-cyclotron waves (ICW), among other modes \cite[see e.g.][]{bale2005,kiyani2012enhanced}. 



In order to explore the relationship between the observation of ion-scale waves and the properties of the inertial range turbulence, we have introduced a quantity that enables the identification of waves and quantitatively assesses their presence in the time series. 
The technique for the identification of the waves is described in detail in~\citet{Yuri2021}, where it has been used to determine that the observed fluctuations are most likely ion cyclotron waves. A brief description of the technique is given in the following.
The first step is to rotate the magnetic field into the field-aligned coordinates using  $\mathbf{B}$ low-pass filtered at 0.01~Hz as the background magnetic field. The power-spectrum of the resulting transverse component, $\delta B_\perp$, is shown in the third panels from the top of Figure~\ref{fig:zoom}.
Subsequently, the coherence between the two perpendicular magnetic field components is computed (fourth panel). 
%
%
This will have large values if circularly-polarized ion-scale waves are present. 
If the coherence is larger than an arbitrary threshold (0.65) in a frequency range near or below the proton gyro-frequency, the phase angle between the two perpendicular magnetic components is also computed (fifth panel), allowing to determine the fluctuations handedness; as we are interested in circularly-polarized waves, we exclude the phases outside the intervals $+90^{\circ}\pm45^{\circ}$ (right-handed waves) and $-90^{\circ}\pm45^{\circ}$ (left-handed waves).
%
%

Using the above indicators, it is therefore possible to unambiguously identify regions with circularly-polarized wave activity both in time and frequency. 
The perpendicular magnetic power is finally integrated in the identified wave intervals and frequency band (with lower and higher frequencies $f_1$ and $f_2$ respectively to be identified according to the above criteria), namely within the wave patches clearly visible in the phase angle plot. 
This procedure provides the time series of one frequency-integrated local parameter $Q_w=\int^{f_1}_{f_2}{\delta B_\perp^2 df}$, defined for each data point in the time series, indicating the total power associated to the wave-like fluctuations. Note that the parameter is not computed outside of the wave patches. 
The bottom panels of Figure~\ref{fig:zoom} shows two examples of wave parameter $Q_w$ for samples 26 and 27.
Interval 26 (left panel) with highly irregular and intermittent behaviour, capturing the corresponding wave patches observed in the scalograms. 
In the adjacent interval 27 (right panel), waves are nearly absent, and accordingly the wave parameter values are negligible.

Finally, using the time series of frequency-integrated wave power, two similar global parameters can be computed to quantitatively assess the occurrence of waves within each interval. The first one is simply obtained as the time-integrated power $\bar{Q}_w=\int{Q_{w}}dt$, the integration being intended over each interval. The second one is the average over the interval $\langle {Q}_w\rangle = \bar{Q}_w/\Delta \mathcal{W}$, taking into account the density of waves within each interval. The obtained values are listed in Table~\ref{table2:H_slope_kurt} for all intervals. In some occasions, when no waves were identified, the parameters were set to 0.
The wave parameters $\bar{Q}_w$ and $\langle {Q}_w\rangle$ will be used in Section ~\ref{sec:discussion} to determine possible correlations with the turbulence parameters.

	\section{Discussion}
    \label{sec:discussion}

Once the turbulent properties of the fluctuations and the presence of ion-scale waves have been quantitatively assessed, it is possible to investigate correlations between the two phenomena. This may help understanding the dynamical processes of solar wind plasmas, and in particular the cross-scale coupling between fluid and sub-ion processes. 
Using the results of the analysis for the 36 samples, correlation coefficients have been computed between pairs of parameters of solar wind ($V_{sw}$ and $\theta_{vb}$), turbulence (Hurst number, spectral exponent, kurtosis and intermittency), and waves (the two estimators presented in Section~\ref{sec:waves}). Both linear (Pearson) and nonlinear (Spearman) coefficients have been computed. For each pair of parameters, the largest of the two has been considered. The complete list of coefficients for all pairs is presented in Table~\ref{tab:corre}. 
As expected, some of the parameters are trivially correlated with each other, such for example those related to different scaling exponents of the same field, or the two wave parameters. 
Others display known correlations, associated with the nature of the solar wind intervals.
However, despite the high variability of the parameters and the experimental conditions, some non-trivially related pairs show moderate, non-negligible correlation. These are highlighted in bold in Table~\ref{tab:corre}. 
The most interesting correlation was found between the wave indicators ($\bar{Q_w}$ and $\langle Q_w\rangle$) and the intermittency parameters ($\mu$ and $\kappa$). 
For example, $C(\langle Q_w\rangle,\mu)=-0.5$ indicates that intervals with substantial presence of ion-scale waves are likely to show reduced intermittency. This observation clearly highlights the link between the characteristics of the fluid-scale turbulent cascade to the excitation of waves at ion scales.

\begin{figure*}
		\includegraphics[scale=0.33]{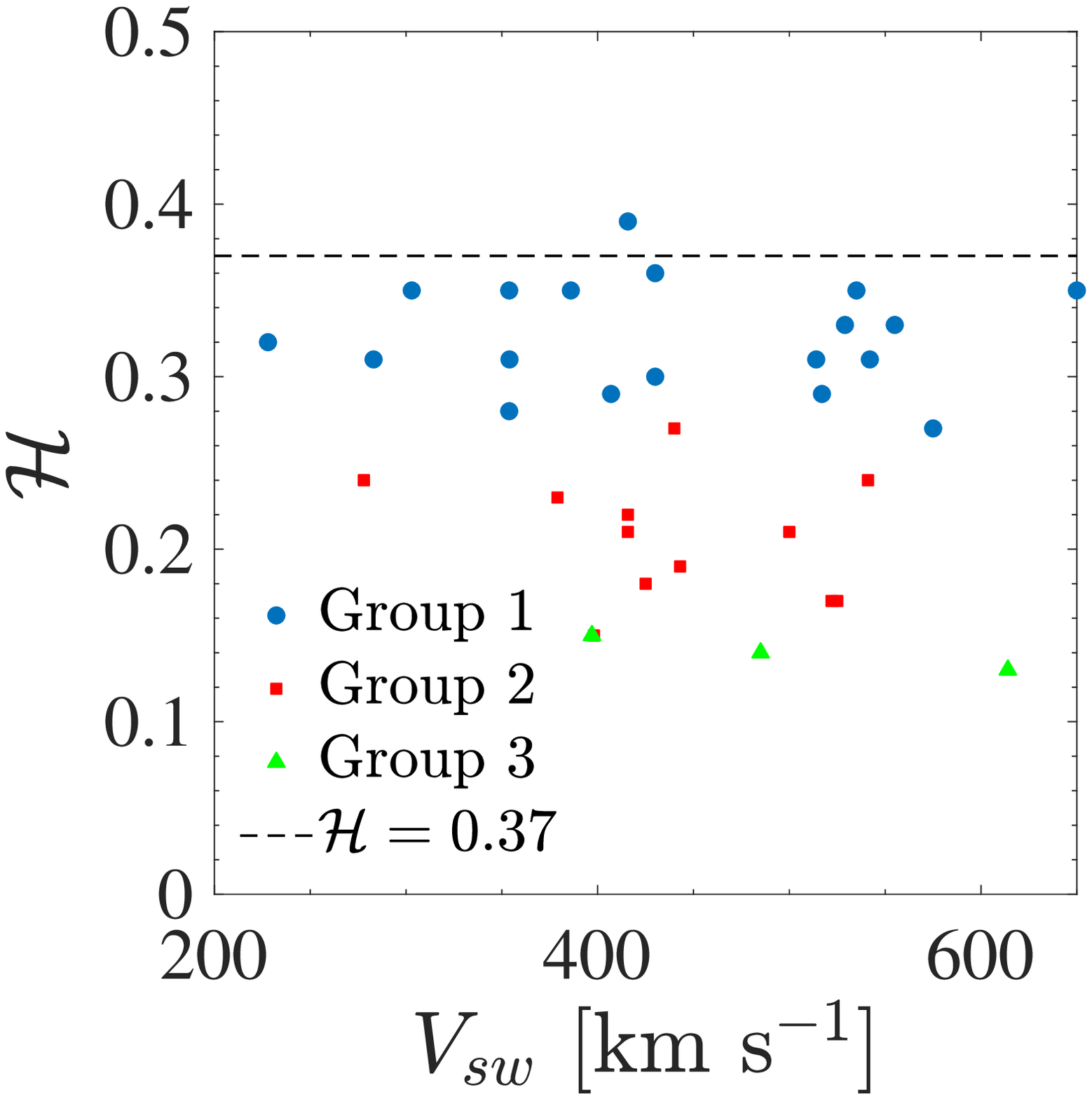}\includegraphics[scale=0.33]{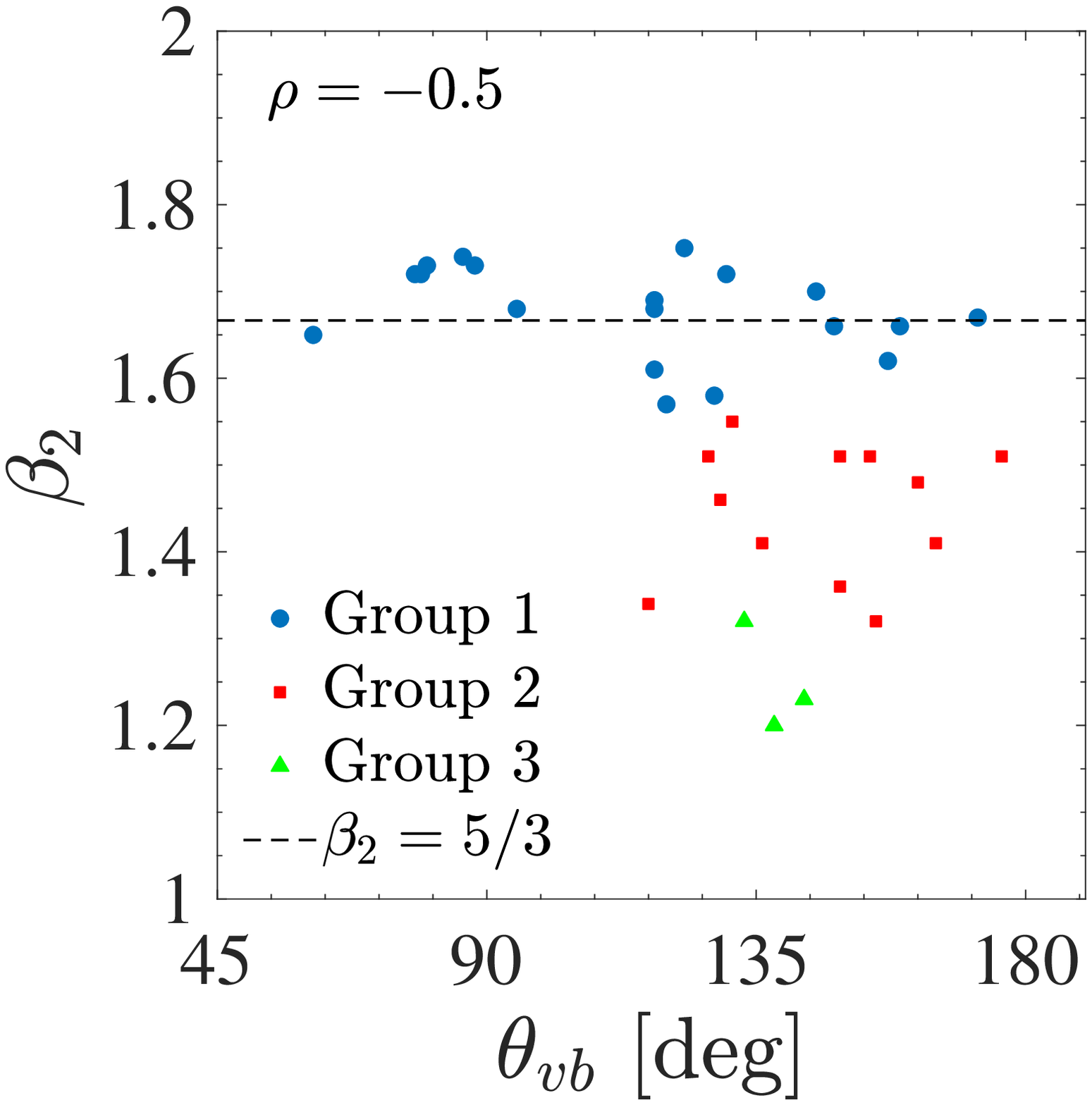}\includegraphics[scale=0.33]{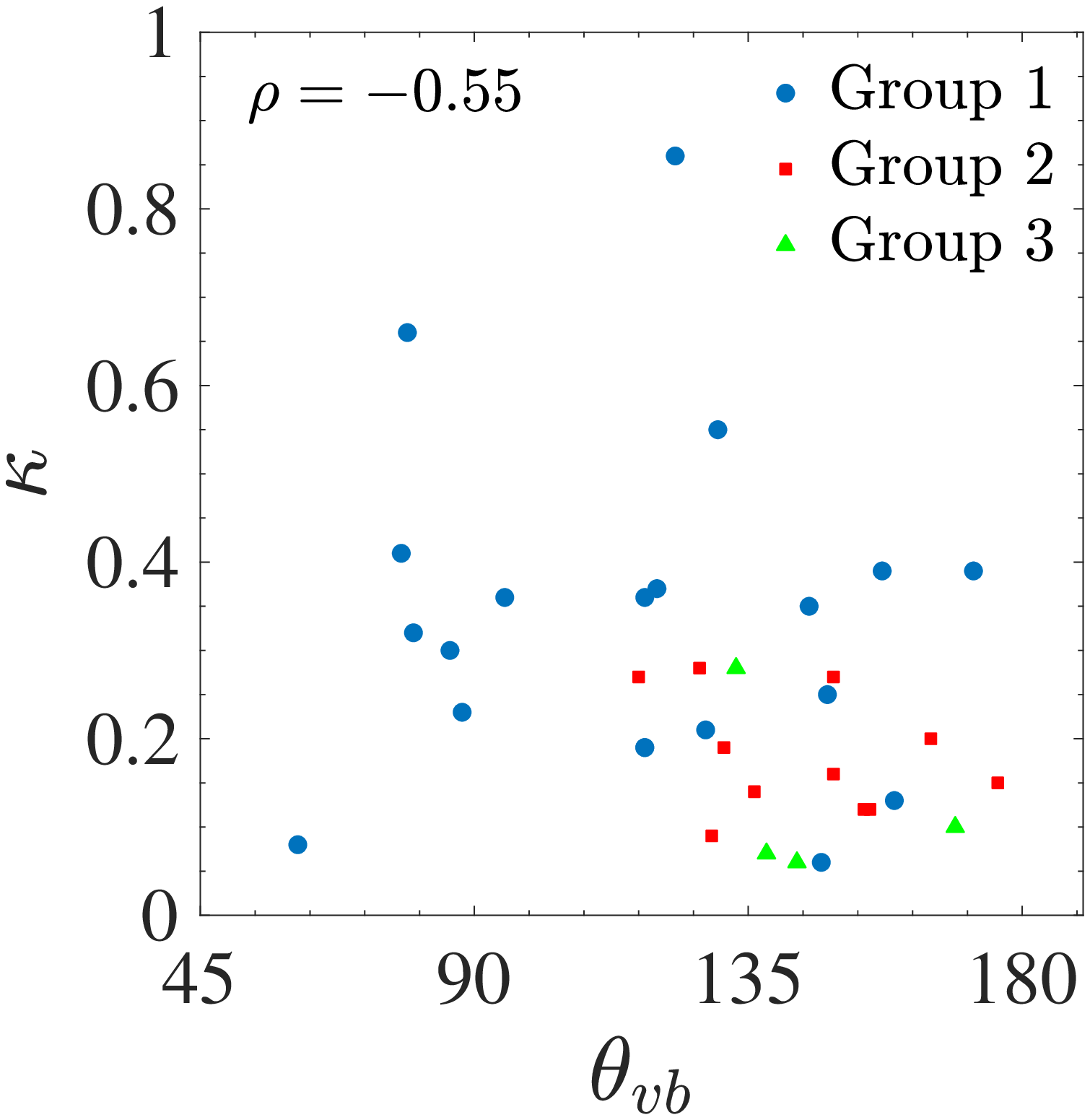}
		\includegraphics[scale=0.33]{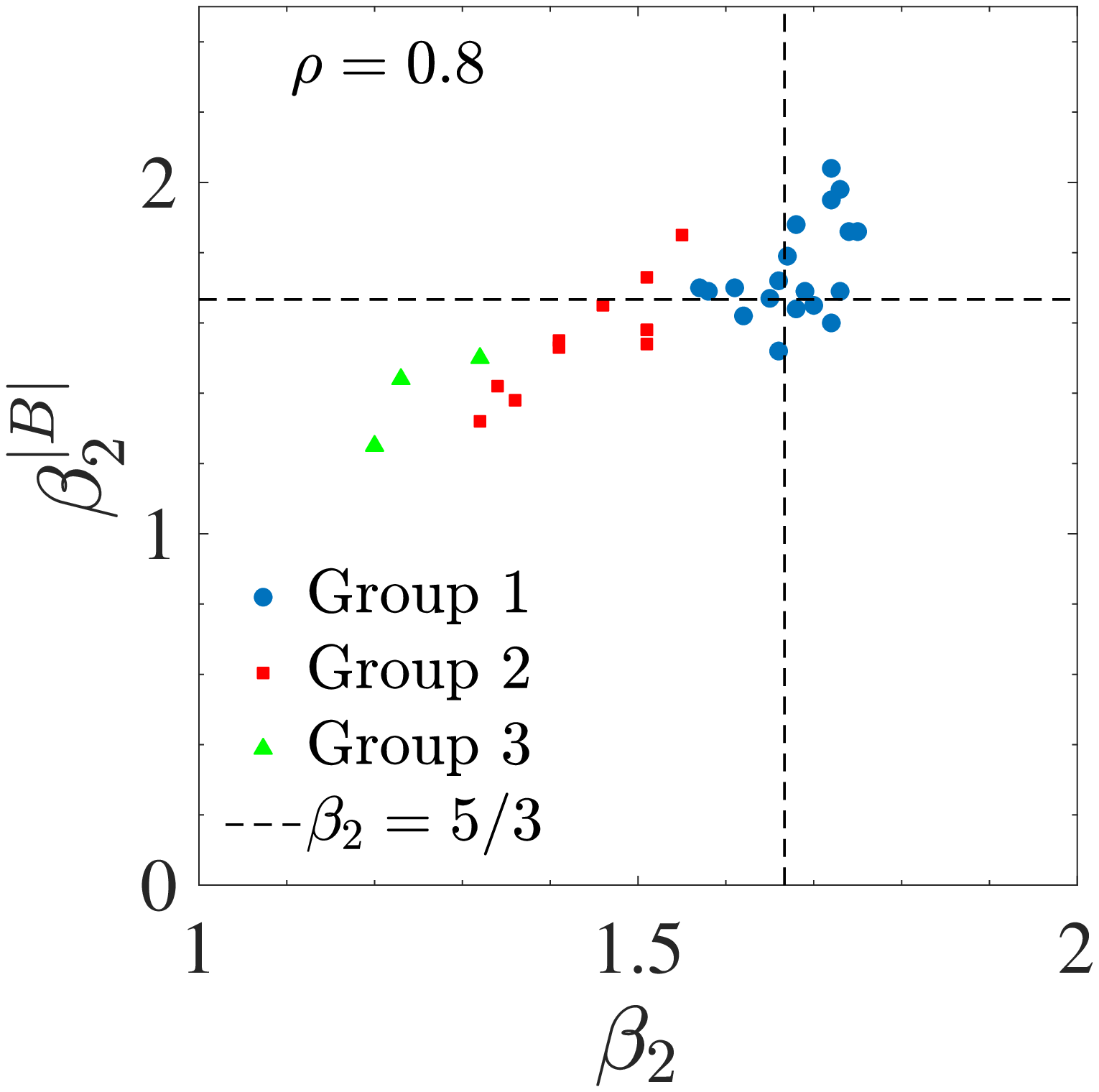}\includegraphics[scale=0.32]{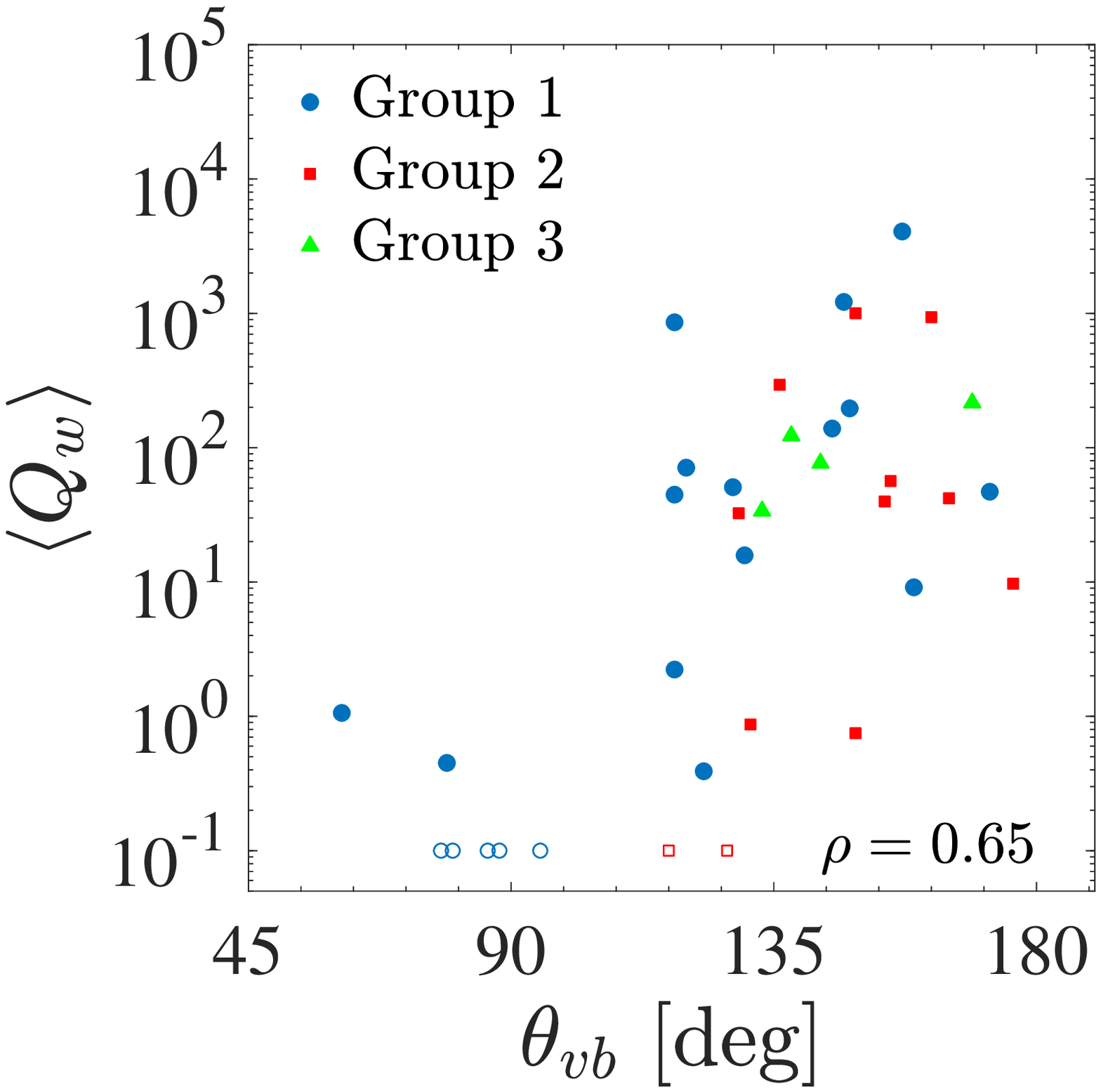}\includegraphics[scale=0.32]{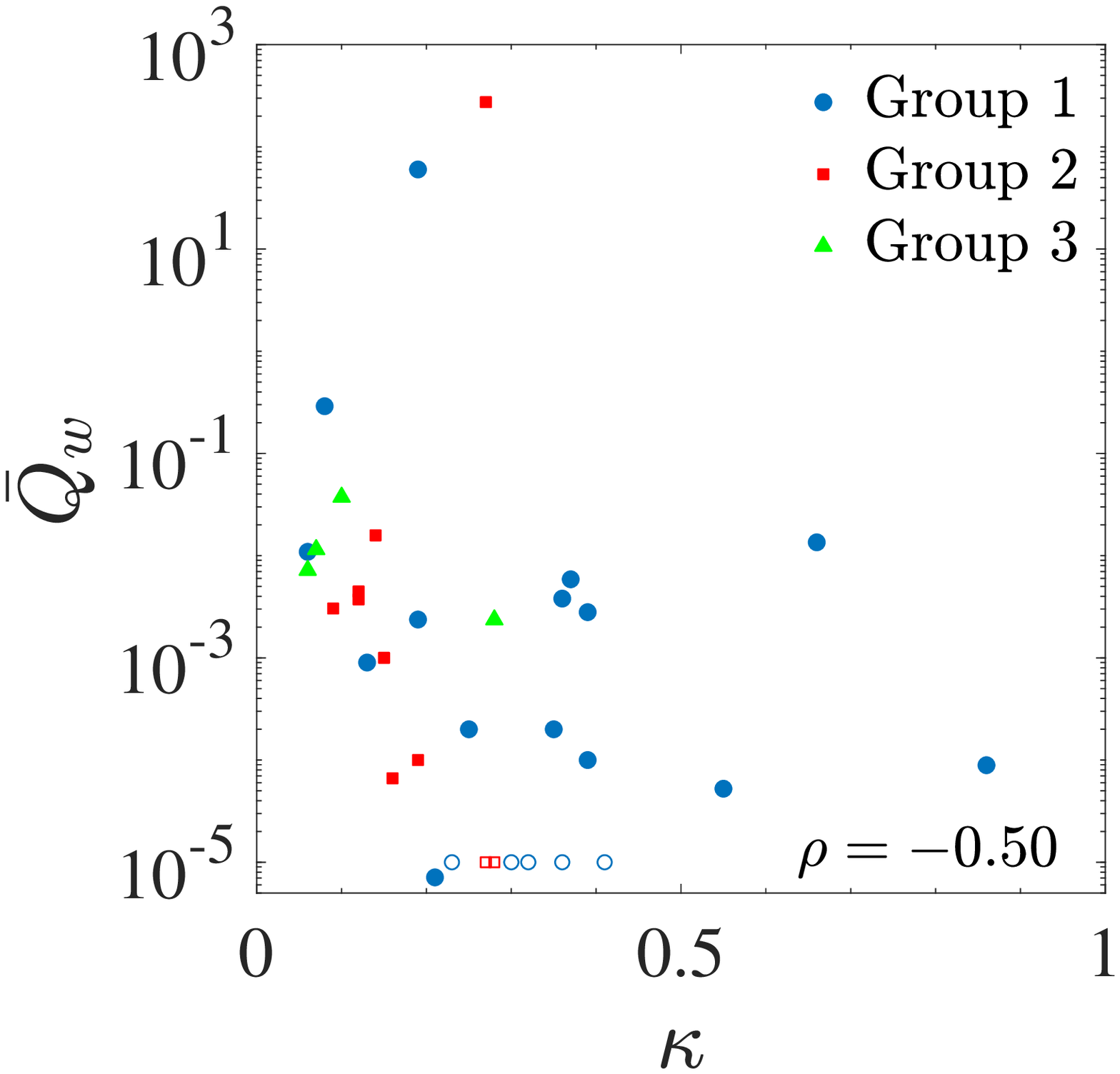}
		\caption{Scatter plots of pairs of solar wind, turbulence and wave parameters  (see Table~\ref{table1:data_params} and~\ref{table2:H_slope_kurt}). Blue, red and green points indicate intervals of group 1, 2 and 3 respectively. Blue open circles in the bottom central and left panels represent intervals with wave zero parameters, for clarity, their value is set to 0.1 in order to be represented in the logarithmic vertical axis.   
		Vertical and horizontal lines indicate standard fluid turbulence reference vales. The correlation coefficient is indicated (Table~\ref{tab:corre}).}
		\label{fig:scatter}
\end{figure*}

A more visual description is provided by the scatter plots of pairs of parameters listed in Tables~\ref{table1:data_params} and~\ref{table2:H_slope_kurt}, shown in Figure~\ref{fig:scatter} for some pairs of parameters from Table~\ref{tab:corre}. 
In all panels, the samples are color-coded according to their group as determined in Section~\ref{sec:hurst} (group 1: blue circles; group 2: red squares; group 3: green triangles). Whenever relevant, vertical or horizontal lines indicate typical value of the parameters for standard fluid turbulence. 
The top-left panel, plotting the wind speed and the density Hurst number, highlights the clear separation between group 1 (mostly large Hurst number) and groups 2 and 3 (smaller Hurst number). Additionally, it clearly shows that while intervals of group 1 belong to both fast and slow wind, nearly all intervals of groups 2 and 3 (with one single exception) belong to faster solar wind. It is worth noting that the flow speed does not necessarily act as an ordering parameter for Alfv\'enicity, though slow wind is generally less Alfv\'enic than fast wind~\citep[see for example][and references therein]{DAmicis2021}. Specifically, fast wind can exhibit different levels of Alfv\'enicity. In this respect, the top-left panel suggests that the lack of Alfv\'enicity in the slow wind assures a more developed turbulence, while possible enhancements in Alfv\'enic nature of the fluctuations in some (though not all) fast wind samples may prevent plasma from fully developing into a turbulent state. 
It turns out that in the samples studied here the fast wind can include both standard and reduced turbulence intervals (e.g. with shallower spectra), depending on the corresponding level of Alfv\'enic fluctuations.
We remind that while solar wind turbulence is likely strongly driven by Alfv\'enic fluctuations, these need to include both counterpropagating modes in order to effectively generate nonlinear interactions. On the other hand, if the fluctuations are unbalanced, with one mode prevailing over the other (typically resulting in a definite sign large cross-helicity), then the sweeping effect strongly reduces the nonlinear interactions, resulting in weaker turbulence and shallower spectra~\citep{Dobrowolny1980}.

The top-center panel of Figure~\ref{fig:scatter} shows that the angle between the magnetic field and the radial direction (approximately corresponding to the velocity vector and in turn to the sampling direction, at such distances) is also relevant to the turbulence. In particular, for group 1 intervals the Kolmogorov spectrum (dashed line) is observed at all angles. On the contrary, groups 2 and 3, characterized by a shallower spectrum, only include intervals with nearly radial field. This suggests that during intervals belonging to groups 2 and 3, Solar Orbiter sampled parallel fluctuations (namely the slab component of turbulence), which are generally less evolved and likely more Alfv\'enic with respect to 2D turbulence. Interestingly, this result is in contrast with critical balance theory~\citep{Goldreich1995,Telloni2019}, which predicts a steeper spectrum (with a scaling close to $-2$) for parallel fluctuations~\citep[an interesting discussion on the validity and relevance of critical balance in solar wind turbulence is provided in][]{Oughton2020}.

The top-right panel of Figure~\ref{fig:scatter} shows that strong intermittency (large kurtosis) is mostly observed in intervals with quasi-perpendicular field. Additionally, it highlights the good correlation existing between the angle and the intermittency exponent $\kappa$, demonstrating that $\theta_{vb}$ is a good ordering parameter for intermittency. As mentioned above, in the studied intervals the solar wind plasma is likely to be more Alfv\'enic at quasi-parallel angles, where the turbulence is only poorly developed. The stochastic nature of the Alfv\'enic fluctuations tends to reduce the intermittency, which is indeed lower at larger angles. On the other hand, at quasi-perpendicular angles, where the turbulence is more fully developed (possibly in association with reduced Alfv\'enic fluctuations), the mitigating effect of Alfv\'enicity is lower and the coherent structures advected by the wind tend to emerge, resulting in the observed increasing intermittency.

The bottom-left panel of Figure~\ref{fig:scatter} shows the strong overall correlation between the spectral exponents of electron density and magnetic field magnitude. 
For the intervals of group 1, spectral exponents of both fields are mostly consistent with the standard Kolmogorov value. 
On the contrary, the evident linear correlation for the more variable exponents of groups 2 and 3 strongly suggests the Alfv\'enic nature of the fluctuations, with well correlated compressive magnetic magnitude and plasma density fluctuations. 
It is indeed worth reminding that density behaves as a passive scalar (it reproduces the magnetic field magnitude characteristics) only in the Alfv\'enic solar wind, where the contribution of compressive fluctuations is negligible. 
In this perspective, for intervals in groups 2 and 3 the plasma density can be considered as a proxy of the magnetic field for the turbulent properties (spectral scaling, intermittency, etc.).

In the bottom-center panel of Figure~\ref{fig:scatter}, the correlation between the angle and the wave density parameter is shown. In this case, no clear separation between the three groups is observed. However, it is evident that for intervals with perpendicular field the wave density is always small (note that 5 intervals of group 1 for which $\langle Q_w\rangle=0$ have been artificially represented on the logarithmic vertical axis by the open blue circles at 0.1). The four intervals of group 3 are also characterized by large presence of waves. This is in good agreement with the expectations. Indeed, the presence of KAWs (at quasi-perpendicular angles) and ICWs (at quasi-parallel angles) strictly depends on the presence of Alfv\'enic fluctuations at fluid scales. Larger Alfv\'enicity is associated with enhanced presence of waves at ion scales, as first shown by~\citet{2015ApJ...811L..17B} (see Fig. 2 therein) on a single case study, and then corroborated on a statistical data set by~\citet{Telloni2019b} (see Fig. 3c therein). 

Finally, the bottom-right panel of Figure~\ref{fig:scatter} shows the correlation between the intermittency scaling exponent $\kappa$ and the total wave power $\langle Q_w\rangle$. For group 1 intervals with no waves, the same representation as in the bottom-center panel has been adopted. The observed correlation is a very interesting result. Indeed, despite the scattered plot, a general trend is evident: stronger intermittency intervals have less wave activity. 
This is in striking agreement with a scenario in which for higher the Alfv\'enic fluctuations (which implies a lower intermittency), the presence of waves at ion scales (and, in turn, the related measured energy) is larger. This has been very recently validated by the statistical work by~\citet{Telloni2019b}.

The overall conclusion gained from the examination of Figure \ref{fig:scatter} is that 2D fluctuations (fluctuations sampled at quasi-perpendicular angles with the magnetic field) are always characterized by strong Kolmogorov turbulence, strong intermittency  and absence of wave activity. A a less Alfv\'enic content is also suggested. As discussed above, all these fluid and kinetic characteristics are strictly related to each other. On the other hand, slab fluctuations (fluctuations sampled at quasi-radial directions) are associated with less developed turbulence (smaller spectral exponent and intermittency parameters) and stronger ion-scale wave activity. A higher Alfv\'enic content (which acts to make less efficient the nonlinear interactions) can be inferred for these intervals. 

In this respect, the speed of the solar wind flows does not seem to be an order parameter, Alfv\'enicity being a more suitable one.

	\section{Conclusions}
    
\label{sec:conclusions}

We have presented the first analysis of turbulence and intermittency of the solar wind electron density measured by the RPW instrument on-board the Solar Orbiter spacecraft. 36 intervals were selected during the month of June 2020, when Solar Orbiter was located in the inner heliosphere, approximately at 0.5 AU from the Sun. 
The study was performed using standard analysis techniques as well as the Empirical Mode Decomposition. 
It was found that the intervals could be separated in three groups, according to their agreement with standard turbulence parameters. The distinction was attributed to the different level of Alfv\'enic fluctuations, which in the absence of plasma measurements was inferred from the turbulence characteristics. Using the magnetic field measurements from the MAG instrument, the presence of ion-scale waves has been detected through coherence analysis. These waves have been identified as being mostly ion cyclotron waves. A new parameter has been introduced to determine quantitatively the energy associated to waves in a given frequency range. Making use of such parameter, two estimators were introduced to assess the overall wave activity within each interval. Comparing these wave parameters with the turbulence indicators, it has been found that intervals with enhanced presence of waves are also characterized by anomalous turbulence and weaker intermittency of the solar wind density. 
While the study of statistical correlations cannot determine the causality relation between the phenomena, it definitely provides constrains and, in this specific case, can help understanding the cross-scale connection between the fluid scales and the ion-scales. The possible role of Alfv\'enic fluctuations in driving both the observed ion cyclotron waves and the reduced turbulence and intermittency was highlighted. 

The preliminary results described in this paper demonstrate the unprecedented high quality of the Solar Orbiter RPW electron density data estimated using the probe-to-spacecraft potential. The excellent performances of the EMD-based analysis allowed the accurate determination of turbulence parameters. The results described here represent the most detailed description of turbulence of solar wind density fluctuations so far.  
While these results are generally in line with previous observations at 1 AU, future studies of Solar Orbiter measurements will finally allow us to determine the radial evolution of the properties of density turbulence. 

Furthermore, the novel parameter providing quantitative assessment of the presence of waves helped identify an important relationship between fluid-scale turbulence and ion-scale phenomena in the solar wind plasma.
The future study of measurements including magnetic field and velocity fluctuations, and more extended statistical analysis could provide a deeper understanding of such relationship.

    \begin{acknowledgements}
    
Solar  Orbiter  is  a  space  mission  of  international  collaboration between ESA and NASA, operated by ESA. We thank the entire Solar Orbiter team and instrument PIs for data access and support. Solar Orbiter data are  available  at  http://soar.esac.esa.int/soar.  
The RPW instrument has been designed and funded by CNES, CNRS, the Paris Observatory, The Swedish National Space Agency, ESA-PRODEX and all the participating institutes.
Swedish co-authors are supported  by the Swedish Research Council, grant 2016-05507, and Swedish National Space Agency grant 20/136. 
CNES and CDPP are acknowledged for the support to the French co-authors.
Solar Orbiter magnetometer operations are funded by the UK Space Agency (grant ST/T001062/1). TSH is supported by STFC grant ST/S000364/1.
FC acknowledges the contribution received from EU-H2020 program ERA-PLANET through the project "Integrated Global Observing Systems for Persistent Pollutants" (iGOSP) (Grant Agreement: 689443), funded under H2020-SC5-15-2015 “Strengthening the European Research Area in the domain of Earth Observation”, from FET Proactive project "Towards new frontiers for distributed environmental monitoring based on an ecosystem of plant seed-like soft robots" (I-Seed), funded under Horizon 2020 research and innovation programme (Grant agreement: 101017940), and from EU-H2020 project "EuroGEO Showcases: Applications Powered by Europe" (e-shape) (Grant Agreement: 820852), funded under H2020-SC5-2018-2 "Strengthening the benefits for Europe of the Global Earth Observation System of Systems (GEOSS) - establishing EuroGEO".
LSV was funded by the Swedish Contingency Agency grant 2016-2102 and by SNSA grant 86/20. 
CLV was partially supported by EPN project PIM-19-01.

    \end{acknowledgements}
	
	\bibliographystyle{aa}
	\bibliography{V24}

\begin{appendix}	\section{Tables}

	\begin{table*}
	\caption{List of turbulence and wave parameters obtained from the analysis of the intervals (\# indicates the interval number): the Hurst numbers $\mathcal{H}$, $\mathcal{H}_{|B|}$ $\mathcal{H}_r$ (for electron density, magnetic field magnitude and magnetic radial component respectively); the Hilbert spectral exponents  $\beta_2$, $\beta_2^{|B|}$ and $\beta_2^r$; the total and averaged wave parameters $\bar{Q}_\mathcal{W}$ and $\langle Q_\mathcal{W}\rangle$; the kurtosis scaling exponent $\kappa$; and the intermittency parameter $\mu$ for the electron density fluctuations. Absent values (---) indicate lack of power-law scaling. For the wave parameters, null values (0) indicate no wave power above the selected threshold.}             
	\label{table2:H_slope_kurt}      
	\centering                          
	\begin{tabular}{c c c c c c c c c c c}        
		\hline\hline                 
		\# & $\mathcal{H}$ & $\mathcal{H}_{|B|}$ & $\mathcal{H}_r$ & $\beta_2$ & $\beta_2^{|B|}$ & $\beta_2^r$ & $\bar{Q}_\mathcal{W}$ & 
		$\langle Q_\mathcal{W}\rangle$ & $\kappa$ & $\mu$  \\    
		\hline 			
		\multicolumn{11}{c}{{\bf 07-June-2020}} \\
		\hline
		1  & 0.32$\pm$0.06 & 0.48$\pm$0.09 & 0.36$\pm$0.07 & 1.73$\pm$0.07 & 1.98$\pm$0.12 & 1.69$\pm$0.18 & 0      & 0     & 0.32 & 0.020 \\
		2  & 0.28$\pm$0.05 & 0.40$\pm$0.05 & 0.35$\pm$0.03 & 1.68$\pm$0.09 & 1.88$\pm$0.08 & 1.71$\pm$0.06 & 0      & 0     & 0.36 & 0.015 \\ 
		3  & 0.35$\pm$0.04 & 0.55$\pm$0.07 & 0.32$\pm$0.05 & 1.72$\pm$0.08 & 2.04$\pm$0.09 & 1.68$\pm$0.20 & 0      & 0     & 0.41 & 0.041 \\	
		4  & 0.35$\pm$0.06 & 0.40$\pm$0.08 & 0.31$\pm$0.04 & 1.74$\pm$0.07 & 1.86$\pm$0.14 & 1.65$\pm$0.08 & 0      & 0     & 0 30 & 0.026 \\
		5  & 0.30$\pm$0.06 & 0.38$\pm$0.09 & 0.22$\pm$0.05 & 1.73$\pm$0.08 & 1.69$\pm$0.13 & 1.47$\pm$0.07 & 0      & 0     & 0.23 & 0.032 \\
		6  & 0.31$\pm$0.13 & 0.31$\pm$0.04 & 0.35$\pm$0.04 & 1.58$\pm$0.20 & 1.69$\pm$0.06 & 1.70$\pm$0.08 & 0.004  & 50.8  & 0.36 & 0.006 \\	
		\hline  
		\multicolumn{11}{c}{{\bf 08-June-2020}} \\
		\hline
		7  & 0.36$\pm$0.08 & 0.53$\pm$0.09 & 0.36$\pm$0.09 & 1.72$\pm$0.10 & 1.95$\pm$0.11 & 1.73$\pm$0.16 & 0.002  & 15.8 & 0.19    & 0.035 \\
		8  & 0.15$\pm$0.08 & 0.24$\pm$0.05 & 0.24$\pm$0.06 & 1.34$\pm$0.09 & 1.42$\pm$0.06 & 1.53$\pm$0.08 & 0        & 0     & 0.27 & ---  \\
		9  & 0.21$\pm$0.09 & 0.29$\pm$0.10 & 0.22$\pm$0.04 & 1.51$\pm$0.09 & 1.58$\pm$0.15 & 1.50$\pm$0.07 & 0        & 0     & 0.28 & 0.008 \\
		10 & 0.21$\pm$0.08 & 0.32$\pm$0.05 & 0.29$\pm$0.06 & 1.46$\pm$0.12 & 1.65$\pm$0.09 & 1.51$\pm$0.07 & 0.003  & 32.5  & 0.09   & 0.025 \\
		11 & 0.39$\pm$0.07 & 0.46$\pm$0.10 & 0.35$\pm$0.05 & 1.75$\pm$0.11 & 1.86$\pm$0.13 & 1.69$\pm$0.11 &$\lesssim$10$^{-4}$& 0.39 & 0.86 & 0.055 \\
		12 & 0.27$\pm$0.04 & 0.42$\pm$0.09 & 0.35$\pm$0.04 & 1.55$\pm$0.05 & 1.85$\pm$0.11 & 1.70$\pm$0.06 & $\lesssim$10$^{-4}$ & 0.87    & 0.19 & 0.042 \\
		\hline  
		\multicolumn{11}{c}{{\bf 09-June-2020}} \\
		\hline
		13 & 0.13$\pm$0.03 & 0.24$\pm$0.04 & 0.37$\pm$0.06 & 1.32$\pm$0.06 & 1.50$\pm$0.04 & 1.72$\pm$0.10 & 0.002   & 33.80  & 0.28 & ---   \\
		14 & 0.31$\pm$0.08 & 0.31$\pm$0.08 & 0.36$\pm$0.04 & 1.62$\pm$0.12 & 1.62$\pm$0.13 & 1.71$\pm$0.08 & $\lesssim$10$^{-4}$   & 4.07  & 0.25 & 0.008  \\
		15 & 0.33$\pm$0.11 & 0.35$\pm$0.06 & 0.34$\pm$0.05 & 1.66$\pm$0.12 & 1.72$\pm$0.07 & 1.68$\pm$0.16 & 0.011   & 196 & 0.06 & 0.013  \\
		16 & 0.14$\pm$0.10 & 0.20$\pm$0.10 & 0.32$\pm$0.11 & 1.23$\pm$0.10 & 1.44$\pm$0.11 & 1.63$\pm$0.07 & 0.007   & 76.8  & 0.06 & ---  \\
		17 & 0.35$\pm$0.08 & 0.39$\pm$0.11 & 0.36$\pm$0.11 & 1.70$\pm$0.11 & 1.65$\pm$0.14 & 1.73$\pm$0.15 & 0.014   & 139 & 0.66 & 0.024 \\
		\hline  
		\multicolumn{11}{c}{{\bf 10-June-2020}} \\
		\hline
		18 & 0.17$\pm$0.07 & 0.17$\pm$0.06 & 0.21$\pm$0.08 & 1.32$\pm$0.08 & 1.32$\pm$0.12 & 1.57$\pm$0.13 & 0.005  & 56.5   & 0.12 & 0.008  \\
		19 & 0.19$\pm$0.07 & 0.28$\pm$0.05 & 0.26$\pm$0.04 & 1.41$\pm$0.09 & 1.55$\pm$0.07 & 1.62$\pm$0.08 & 0.016  & 294  & 0.14 & 0.012  \\
		20 & 0.18$\pm$0.07 & 0.18$\pm$0.07 & 0.37$\pm$0.06 & 1.36$\pm$0.05 & 1.38$\pm$0.11 & 1.73$\pm$0.08 & $\lesssim$10$^{-5}$ & 0.75  & 0.16 & -- \\
		\hline  
		\multicolumn{11}{c}{{\bf 11-June-2020}} \\
		\hline
		21 & 0.24$\pm$0.04 & 0.12$\pm$0.04 & 0.35$\pm$0.06 & 1.51$\pm$0.06 & 1.58$\pm$0.06 & 1.73$\pm$0.06 & 0.004  & 39.8  & 0.12  & 0.022   \\
		22 & 0.33$\pm$0.12 & 0.30$\pm$0.07 & 0.26$\pm$0.09 & 1.66$\pm$0.10 & 1.52$\pm$0.10 & 1.51$\pm$0.13 & 0.001 & 9.13   & 0.13   & 0.018   \\
		23 & 0.15$\pm$0.15 & 0.11$\pm$0.08 & 0.33$\pm$0.10 & 1.20$\pm$0.22 & 1.25$\pm$0.11 & 1.64$\pm$0.15 & 0.012   & 122 & 0.07 & ---     \\
		24 & 0.29$\pm$0.10 & 0.34$\pm$0.14 & 0.28$\pm$0.04 & 1.68$\pm$0.18 & 1.64$\pm$0.18 & 1.61$\pm$0.05 & 0.003   & 44.7  & 0.39 & 0.037  \\
		\hline  
		\multicolumn{11}{c}{{\bf 14-June-2020}} \\
		\hline
		25 & 0.22$\pm$0.12 & 0.25$\pm$0.17 & 0.36$\pm$0.09 & 1.51$\pm$0.15 & 1.73$\pm$0.20 & 1.70$\pm$0.15 & 0.27 & 1002  & 0.27 & 0.003 \\
		\hline  
		\multicolumn{11}{c}{{\bf 20-June-2020}} \\
		\hline	
		26 & 0.27$\pm$0.06 & 0.38$\pm$0.09 & 0.28$\pm$0.07 & 1.67$\pm$0.09 & 1.79$\pm$0.11 & 1.56$\pm$0.09 & 0.006 & 47	 & 0.37 & 0.020   \\
		27 & 0.17$\pm$0.04 & 0.24$\pm$0.03 & 0.33$\pm$0.06 & 1.41$\pm$0.06 & 1.53$\pm$0.15 & 1.65$\pm$0.08 & $\simeq$10$^{-6}$ & 0.042 & 0.20 & 0.005 \\
		\hline  
		\multicolumn{11}{c}{{\bf 22-June-2020}} \\
		\hline	
		28 & 0.31$\pm$0.09 & 0.36$\pm$0.06 & 0.32$\pm$0.07 & 1.61$\pm$0.11 & 1.70$\pm$0.09 & 1.70$\pm$0.09 & 60.1  & 858 & 0.19 & --  \\
		29 & 0.29$\pm$0.08 & 0.30$\pm$0.08 & 0.36$\pm$0.04 & 1.57$\pm$0.11 & 1.70$\pm$0.13 & 1.64$\pm$0.09 & $\lesssim$10$^{-5}$ & 0.07  & 0.21 & --  \\
		\hline  
		\multicolumn{11}{c}{{\bf 24-June-2020}} \\
		\hline	
		30 & 0.23$\pm$0.14 & 0.25$\pm$0.05 & 0.27$\pm$0.04 & 1.51$\pm$0.13 & 1.54$\pm$0.09 & 1.54$\pm$0.08 & 0.002  & 9.7    & 0.15 & ---    \\
		31 & ---           & 0.17$\pm$0.15 & 0.21$\pm$0.09 & ---           & 1.31$\pm$0.20 & 1.46$\pm$0.14 & 0.037  & 216  & 0.10 & ---   \\ 
		32 & ---           & 0.25$\pm$0.06 & 0.23$\pm$0.17 & ---           & 1.56$\pm$0.09 & 1.50$\pm$0.20 & 0.29    & 1217 & 0.08 & 0.037   \\
		\hline  
		\multicolumn{11}{c}{{\bf 27-June-2020}} \\
		\hline	
		33 & 0.31$\pm$0.07 & 0.25$\pm$0.07 & 0.31$\pm$0.09 & 1.72$\pm$0.10 & 1.60$\pm$0.09 & 1.68$\pm$0.10 & $\lesssim$10$^{-4}$ & 0.45 & 0.55 & 0.034  \\
		34 & 0.35$\pm$0.06 & 0.40$\pm$0.05 & 0.34$\pm$0.08 & 1.65$\pm$0.06 & 1.67$\pm$0.08 & 1.64$\pm$0.19 & $\lesssim$10$^{-4}$ & 1.06 & 0.39 & 0.039   \\
		35 & 0.35$\pm$0.14 & 0.34$\pm$0.04 & 0.35$\pm$0.03 & 1.69$\pm$0.19 & 1.69$\pm$0.05 & 1.72$\pm$0.05 & $\lesssim$10$^{-4}$ & 2.23 & 0.35 & 0.019  \\	
		\hline  
		\multicolumn{11}{c}{{\bf 29-June-2020}} \\
		\hline	
		36 & 0.24$\pm$0.12 & --- & 0.27$\pm$0.10 & 1.48$\pm$0.18 & --- & 1.50$\pm$0.21 & 0.156 & 938 & --- & --- \\
		\hline
		\hline
			\end{tabular}
		\end{table*}


	\begin{table*}
	\caption{Correlation coefficients between pairs of parameters, including the solar wind speed  $V_{sw}$, the angle between the mean magnetic field and the radial direction $\theta_{vb}$, and the turbulence and wave parameters listed in Table~\ref{table2:H_slope_kurt}. For each pair, the maximum between the linear (Pearson) and nonlinear (Spearman) coefficients is given. For parameters which are not trivially related, non-negligible correlation values are highlighted in bold.}             
	\label{tab:corre}      
	\centering                          
	\begin{tabular}{c|c c c c c c c c c c c c}        
		\hline\hline                 
 Parameter	&	$V_{sw}$ & $\theta_{vb}$ & $\mathcal{H}$ & $\mathcal{H}_{|B|}$ & $\mathcal{H}_r$ & $\beta_2$ & $\beta_2^{|B|}$ & $\beta_2^r$ & $\bar{Q}_\mathcal{W}$ & $\langle Q_\mathcal{W}\rangle$ & $\kappa$ & $\mu$ \\    
		\hline 			
$V_{sw}$            & 1  & 0.28 & -0.13 & -0.13 & 0.11 & -0.14 & -0.09 & 0.18 & 0.15 & -0.24 & -0.20 & -0.28 \\
		\hline
$\theta_{vb}$       &    & 1 & {\bf -0.43} & {\bf -0.54} & -0.18 & {\bf -0.50} & {\bf -0.51} & -0.22 & {\bf 0.59} & {\bf 0.65} & {\bf -0.55} & -0.38  \\ 
		\hline
$\mathcal{H}$       &    &   & 1 & 0.79 & 27 & 0.94 & 0.73 &0.29 & -0.19 & -0.31 & 0.55 & 0.45   \\
		\hline
$\mathcal{H}_{|B|}$ &    &   &   & 1 & 0.27 & 0.82 & 0.91 & 0.24 & {\bf -0.45} & {\bf -0.43} & 0.55 & 0.53 \\
		\hline
$\mathcal{H}_r$     &    &   &   &   & 1 & 0.16 & 0.41 & 0.90 & -0.10 &  0.16 & 0.29 & -0.16 \\
		\hline
$\beta_2$           &    &   &   &   &   & 1 & 0.80 & 0.19 & {\bf-0.45} & {\bf -0.48} & -0.63 & 0.61 \\	
		\hline
$\beta_2^{|B|}$     &    &   &   &   &   &   & 1 & 0.39 & {\bf-0.43} & -0.34 & 0.51 & 0.36 \\  
		\hline
$\beta_2^r$         &    &   &   &   &   &   &   & 1 & -0.07 & -0.11 & -0.34 & -0.14\\
		\hline
$\bar{Q}_\mathcal{W}$ 	&    &   &   &   &   &   &   &   & 1 & 0.92 & {\bf -0.50} & -0.31 \\
		\hline
$\langle Q_\mathcal{W}\rangle$  &    &   &   &   &   &   &   &   &   & 1 & {\bf -0.45} & {\bf -0.50} \\
		\hline
$\kappa$            &    &   &   &   &   &   &   &   &   &   & 1 & 0.42  \\
        \hline
$\mu$                 &    &   &   &   &   &   &   &   &   &   &   &  1    \\
		\hline
		\hline
			\end{tabular}
		\end{table*}

\end{appendix}

\end{document}